\documentclass[aps,prb,amsmath,amssymb,showpacs,twocolumn,superscriptaddress]{revtex4-1}
\usepackage{amsmath}
\usepackage{amssymb}
\usepackage{amsthm}
\usepackage{setspace}
\usepackage{graphicx}
\usepackage{braket}
\usepackage{mathrsfs}
\usepackage[none]{hyphenat}
\usepackage{float}
\usepackage[colorlinks = true,linkcolor = blue,urlcolor  = blue,citecolor = blue,anchorcolor = blue]{hyperref}
\usepackage[utf8]{inputenc}
\usepackage[english]{babel}
\usepackage{bm}
\usepackage{csquotes}
\usepackage{mathtools,xparse}
\usepackage{xcolor} 

\usepackage[normalem]{ulem}

\begin{document}

\title{Semiclassical theory of frequency dependent linear magneto-optical transport in  Weyl semimetals}

\author{Azaz Ahmad}
\affiliation{Department of Physics, BITS Pilani-Hydrabad Campus, Telangana-500078, India.}
\affiliation{School of Physical Sciences, Indian Institute of Technology Mandi, Mandi-175005, India.}
\affiliation{Department of Physics, Indian Institute of Technology Bombay, Powai, Mumbai 400076, India.}
\author{Pankaj Bhalla}
\affiliation{Department of Physics, SRM University, Andhra Pradesh- 522502, India.}
\author{Snehasish Nandy}
\affiliation{Department of Physics, National Institute of Technology, Assam-788010, India.}
\author{Tanay Nag}
\affiliation{Department of Physics, BITS Pilani-Hydrabad Campus, Telangana-500078, India.}

\begin{abstract}

We develop a semiclassical Boltzmann theory for frequency-dependent magneto-optical transport in Weyl semimetals (WSMs), incorporating momentum-dependent relaxation via scattering matrix approach. The interplay of orbital magnetic moment, tilting of Weyl cone, and intervalley scattering is incorporated to obtain the full optical conductivity tensor in the presence of a static magnetic field. We show that, for untilted WSMs with orbital magnetic moment, strong intervalley scattering in the weak ac regime induces a sign reversal of the longitudinal magneto-optical conductivity (LMOC), thereby suppressing the chiral anomaly. In contrast, in the strong ac regime, intervalley scattering fails to neutralize the chiral imbalance within a driving cycle, and no sign reversal is observed. Orbital magnetic moment induces linear magnetic-field contributions, while chiral anomaly yields quadratic response accompanied by expected angular profiles. Tilt direction and orientation strongly affect LMOC such as, transverse tilt gives symmetric non-monotonic behavior, whereas parallel tilt leads to asymmetric, nearly monotonic response. Notably, negative LMOC arises intrinsically for parallel tilt, but requires orbital magnetic moment for transverse tilt. These results highlight frequency-dependent conductivity as a sensitive probe of chiral relaxation in MHz–THz magneto-optical experiments.

\end{abstract}


\maketitle

\section{Introduction}
\label{sec:introduction}


Topological semimetals have emerged as a central platform for exploring electronic
transport phenomena governed by band topology and quantum geometry. Among them,
Weyl semimetals (WSMs) host low-energy quasiparticles described by Weyl fermions, which
arise near isolated band-touching points in momentum space known as Weyl
nodes when the time-reversal and/or inversion symmetries are broken~\cite{bevan1997momentum,wan2011topological,Yan_2017,armitage2018weyl,burkov2011weyl}.
Each Weyl cone carries a definite chirality and acts as a monopole or antimonopole of Berry
curvature, leading to unconventional electromagnetic responses with no analogue
in ordinary metals~\cite{adler1969axial,nielsen1981no,son2012berry,goswami2015optical}.
These topological characteristics manifest prominently in linear-response
transport phenomena such as the anomalous Hall effect, chiral magnetic effect,
and the negative longitudinal magnetoresistance associated with the chiral
anomaly~\cite{burkov2014anomalous,son2013chiral,kim2014boltzmann,sharma2016nernst}.
Experimental observations in Dirac and Weyl materials have firmly established the
role of Berry curvature in shaping their electromagnetic
response~\cite{zhang2016linear,armitage2018weyl}.

Theoretical descriptions of transport in WSMs are commonly formulated
within the linear-response regime, where the induced current is proportional to
the applied electric field. In the presence of a static magnetic field, linear
magnetotransport is profoundly modified by Berry curvature contributions to 
the semiclassical equations of motion, the phase-space measure, and the chiral
charge dynamics. Within the semiclassical Boltzmann transport framework, these ingredients
naturally account for anomaly-induced effects such as positive longitudinal
magnetoconductivity when the electric field $\mathbf{E}$ and magnetic field $\mathbf{B}$ are aligned such that
$\mathbf{E}\!\cdot\!\mathbf{B}\neq0$~\cite{son2013chiral,burkov2014anomalous}.
Importantly, charge conservation imposes nontrivial constraints on the coupled
kinetic equations for Weyl cones of opposite chirality and plays a decisive role
in determining the linear as well as non-linear transport coefficients~\cite{
PhysRevB.107.245141,PhysRevB.102.014307,PhysRevB.104.245122,PhysRevB.103.144308,nag2021magneto,PhysRevB.104.115420,PhysRevB.105.214307,PhysRevB.107.L081110,sadhukhan2025orbital}, particularly in the
presence of intervalley scattering \cite{sharma2023decoupling,ahmad2021longitudinal,ahmad2024geometry,ahmad2025chiral,ahmad2025longitudinal,ahmad2025chiral,varma2024magnetotransport,sharma2017chiral,varma2026chiral,varma2026strain}.

In addition to the Berry curvature, the orbital magnetic moment (OMM) of Bloch electrons
constitutes an essential ingredient of the linear electromagnetic response of
WSMs, as it couples directly with the external magnetic fields and produces
field-dependent corrections to the band energy, quasiparticle velocity, and
density of states. These corrections are particularly important in
magneto-transport and magneto-optical phenomena, where they can qualitatively
modify conductivity and optical absorption spectra~\cite{xiao2010berry,sodemann2015quantum}.

In realistic Weyl materials, these effects are further intertwined with the
presence of a generic tilt in the Weyl dispersion, which could break emergent Lorentz
invariance (for type-II WSMs) and leads to strong anisotropies in both transport and optical
responses. Band tilt significantly modifies the interplay among Berry curvature, OMM, and scattering mechanisms, and may either enhance or suppress magneto-optical responses depending on the frequency and orientation of the applied fields~\cite{soluyanov2015type,ahmad2021longitudinal,das2019linear}. Despite their evident importance, a comprehensive framework for frequency-dependent linear magneto-optical transport, which consistently accounts for Berry curvature, OMMs, generic band tilt, momentum-dependent relaxation times, and charge conservation, remains absent~\cite{gupta2024magneto, Ghosh_2023,ghosh2025topological,nandy2022nonreciprocal,dey2020dynamic}.

In the context of magneto-optical response, it is important to clearly distinguish between the relative strengths of the magnetic field associated with an electromagnetic radiation and an externally applied static Zeeman field. For an electromagnetic radiation, the
magnetic component is substantially 
weaker than a
static magnetic field that is routinely accessible in experiments. Importantly, this static magnetic field Berry curvature- and orbital moment-induced
magnetotransport effects, including chiral anomaly-related responses. Accordingly,
throughout the current study, the magnetic field \(\mathbf{B}\), as it appears in the semiclassical equations of motion and the Boltzmann transport equation, is considered an independent static control parameter, while the electric field \(\mathbf{E}(t)\)
represents the time-dependent driving field associated with electromagnetic
radiation. This separation of roles is standard in semiclassical treatments of
magnetotransport and is essential for a consistent analysis of linear
magneto-optical response~\cite{gao2022suppression,gupta2024magneto}.\\

\begin{figure}
    \centering
    \includegraphics[width=.95\columnwidth]{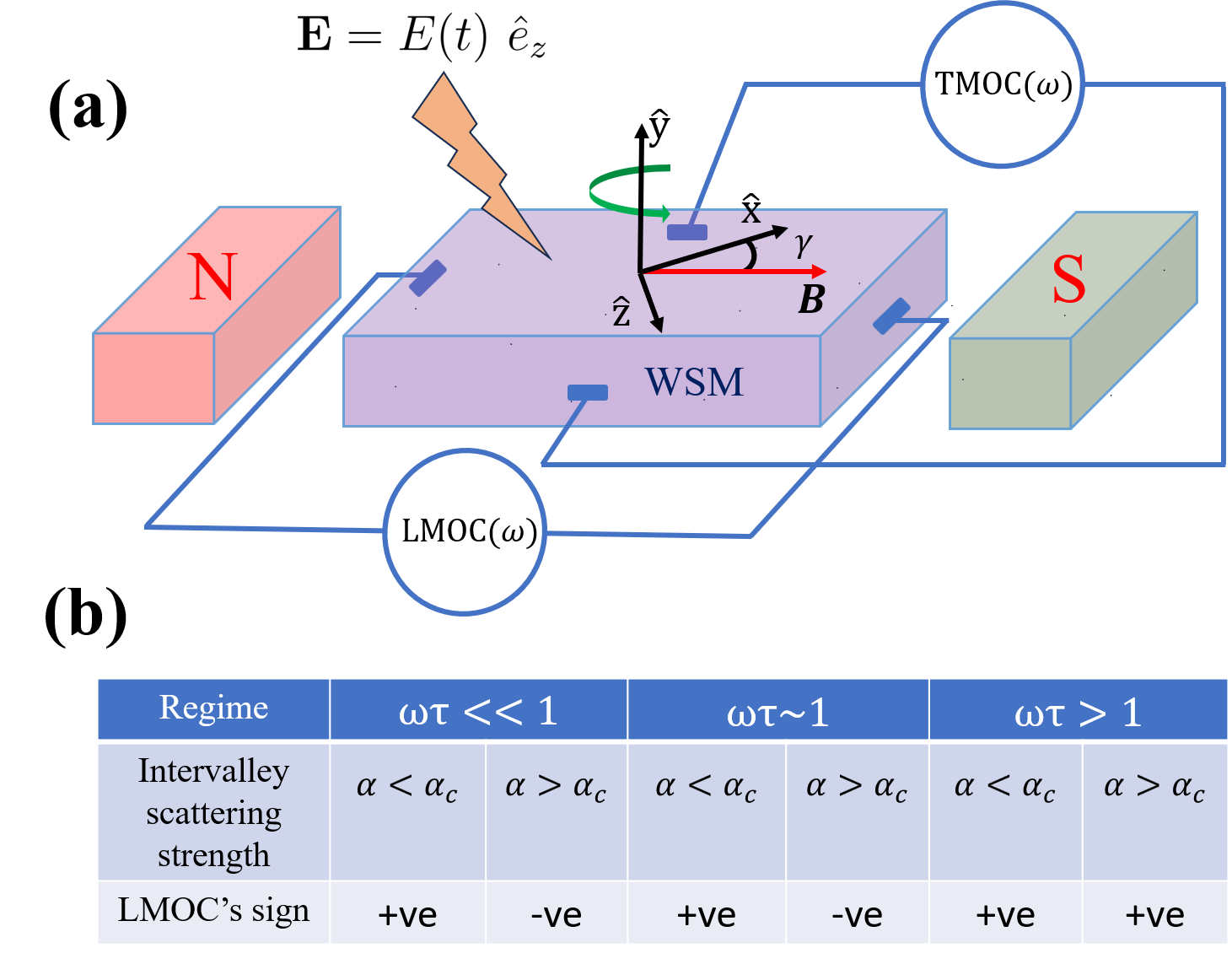}
    \caption{(a) A possible experimental setup to test the angular dependence of longitudinal magneto-optical conductivity  (LMOC) and transverse magneto-optical conductivity (TMOC) under an application of electromagnetic radiation on the WSMs which is placed in a constant magnetic field achieved by two poles of magnets. The WSMs slab can be rotated about $y$-axis to test the angular dependence. The frequency as well as the polarisation direction of the radiation can be tuned externally. (b) We show the sign of LMOC for three different regimes of the electromagnetic radiation and strength of the intervalley scattering.}
    \label{fig:Exp_setup.png}
\end{figure}

We present a semiclassical theory of frequency-dependent linear magneto-optical transport in WSMs within the Boltzmann framework by invoking the scattering matrix formalism that deals with momentum-dependent relaxation time. Our work considers the complex interplay between the OMMs, the generic tilt of the Weyl dispersion, intervalley scattering and the electromagnetic radiation while deriving the full frequency-dependent conductivity tensor in the presence of static external magnetic field.  For untilted Weyl cones, increasing the driving frequency suppresses the chiral imbalance and weakens the impact of intervalley relaxation. As a result, the sign reversal of the longitudinal magneto-optical conductivity (LMOC) is observed in the weak ac limit at strong intervalley scattering while sign reversal is absent in the strong ac limit \cite{sharma2023decoupling,ahmad2021longitudinal,ahmad2024geometry,ahmad2025chiral,ahmad2025longitudinal,ahmad2025chiral,varma2024magnetotransport,sharma2017chiral,varma2026chiral,varma2026strain,ahmad2026magnetotransport}.  {The OMM introduces linear as well as higher-order magnetic-field corrections to the LMOC and transverse magneto-optical conductivity (TMOC) through the modified dispersion, velocity, Fermi surface, and phase-space factor, while a leading quadratic response with the usual angular profile originates from the chiral anomaly term $(\mathbf E \cdot \mathbf B)\mathbf B$.}
The direction of the tilt in the Weyl cone, which can be parallel or transverse with respect to the magnetic field and the relative orientation of the tilt between the pair of Weyl cones, which can be identical or opposite,  play significant roles in determining the response. The LMOC exhibits a symmetric (asymmetric) but non-monotonic (almost monotonic) dependence on tilt strength for transversely (parallelly) tilted Weyl cones which are oriented oppositely.  Remarkably, the above response for parallelly tilted Weyl cones
is accompanied by a negative  LMOC  without OMM which happens to be the key ingredient to realize negative LMOC for the transversely tilted case. On the other hand, Weyl cones with identical tilt orientation exhibit qualitatively similar responses for both parallel as well as transverse tilt directions. 
All the above findings under linearly polarized light remain qualitatively unchanged for circularly polarized light, thus validating our theory for a general irradiation.
Therefore, our results establish frequency-dependent conductivities as a sensitive probe of chiral charge relaxation for magneto-optical experiments in the MHz–THz regime, as depicted in Fig.~\ref{fig:Exp_setup.png}.


The article is organized as follows. In Sec.~\ref {Sec:Model}, we introduce an effective low-energy minimal model of a generic WSMs which includes a tilt term. In Sec.~\ref {sec:MBT}, we discuss the quasi-classical Boltzmann transport formalism incorporating momentum-dependent scattering along with charge conservation at finite frequency electromagnetic radiation to calculate the different electromagnetic responses and related mathematical parts are discussed in the Appendices. In Sec.~\ref{r_results}, we present our results. Finally, we conclude our findings in Sec.~\ref{sec:conclusion}.

\section{Low-Energy Effective Hamiltonian}
\label{Sec:Model}


We commence our analysis by establishing the low-energy effective Hamiltonian that describes the quasiparticle excitations in the vicinity of the Weyl cones. For a WSM incorporating tilt anisotropy, the Hamiltonian is expressed as~\cite{yan2017topological, ahmad2024geometry,das2019linear,ahmad2025chiral,lv2015experimental}:
\begin{align}
H_{\mathrm{WSM}}(\mathbf{k})&= \textcolor{black}{\sum_{\chi=\pm1}H^{\chi}_{\mathbf{k}}}\nonumber\\&\equiv
\sum_{\chi=\pm1} \bigg[ \chi\hbar v_{F}\mathbf{k}\cdot\boldsymbol{\sigma} + \hbar v_{F}(t_z^{\chi} k_z + t_x^{\chi} k_x )\bigg],
\label{Hamiltonian}
\end{align}
where $\chi = \pm 1$ denotes the chirality (also known as topological charge) of the Weyl cone, and $v_{F}$ represents the Fermi velocity. The wave vector $\mathbf{k}=(k_x,k_y,k_z)$ is defined relative to the Weyl cone, while $\boldsymbol{\sigma}=(\sigma_x,\sigma_y,\sigma_z)$ acts on the pseudospin subspace. The parameters $t^{\chi}_{x,z}$ govern the geometric tilt of the Weyl cones along $\hat{x}, ~\hat{z}$ directions, respectively. The orientation and degree of anisotropy of the cones are determined by the magnitude and sign of these tilt parameters, respectively. In the context of inversion symmetry-broken Weyl semimetals, one obtains a minimum of two pairs of nodes. Note that the chiralities  are assigned to ensure the global violation of inversion symmetry \cite{ahmad2021longitudinal,ahmad2023longitudinal,mandal2022chiral,das2019linear,das2022nonlinear,nielsen1981no}. We restrict our calculation to a single pair of Weyl cones, noting that the inclusion of additional pairs only leads to quantitative modifications of the total 
response~\cite{li2021nonlinear,ahmad2025chiral}. The energy dispersion relation \textcolor{black}{associated with a particular cone with chirality $\chi$} is given by:
\begin{align}
\epsilon^{\chi, \pm}_{\mathbf k}=\pm{\hbar v_{F}}|\mathbf{k}| + \hbar v_{F}(t^{\chi}_z k_z+t^{\chi}_x k_x),
\label{Dispersion}
\end{align}
where $+$ and $-$ signs represent conduction and valence bands, respectively and $|\mathbf k|=\sqrt{k_x^2+k_y^2+k_z^2}$. Note that  $\epsilon^{+1, +}_{\mathbf k}=\epsilon^{-1, -}_{\mathbf k}$ and  $\epsilon^{-1, +}_{\mathbf k}=\epsilon^{+1, -}_{\mathbf k}$.  In this work, we restrict the Fermi energy to be in the conduction band and consequently, dropped the band index for simplicity. To be precise, $\epsilon^{\chi, +}_{\mathbf k}\equiv \epsilon^{\chi}_{\mathbf k}$.  The OMM, arising from the self-rotation of the Bloch wave packet, plays an important role in the semiclassical dynamics. The OMM for the conduction band is given by~\cite{xiao2010berry}:
\begin{align}
\mathbf{m}^{\chi}_{\mathbf k}= -\frac{ie}{2\hbar} \mathrm{Im} 
\left\langle \nabla_{\mathbf{k}} u^{\chi}_{ \mathbf{k}} \left| 
\left[ \epsilon^{\chi}_{\mathbf k} - \hat{H}^{\chi}(\mathbf{k}) \right] 
\right| \nabla_{\mathbf{k}} u^{\chi}_{\mathbf{k}} \right\rangle,
\label{Eq:OMM_formula}
\end{align}
\textcolor{black}{with $| u^{\chi}_{\mathbf{k}} \rangle$ is the periodic part of the Bloch wave function}. The coupling of the OMM to an external magnetic field $\mathbf{B}$ modifies the band dispersion according to: \textcolor{black}{$\tilde{\epsilon}^{\chi}_{\mathbf{k}}=\epsilon_{\mathbf{k}}^{\chi}-\mathbf{m}^{\chi}_{\mathbf{k}}\cdot\mathbf{B}$.}  We, henceforth, refer  to $\tilde{\epsilon}^{\chi}_{\mathbf{k}}$ as 
${\epsilon}^{\chi}_{\mathbf{k}}$ for ease of notation. 
Considering spherical polar coordinate system $\mathbf k=|\mathbf{k}|(\sin \theta \cos \phi,\sin \theta \sin \phi, \cos\theta)$, with $\theta$ and $\phi$ as polar and azimuthal angles, respectively, the constant energy contour at the Fermi energy $\epsilon_F$ is obtained as~\cite{ahmad2021longitudinal,sharma2020sign,varma2026chiral}:
\begin{align}
k^{\chi}_{F} = \frac{\epsilon_{F}+\sqrt{\epsilon^{2}_{F} - {\color{black} \chi e v_{F} B \eta^{\chi}_{\theta\phi}} {\color{black}\beta_{\theta \phi \gamma}}}}{{\color{black}\eta^{\chi}_{\theta\phi}}},
\end{align}
where the angular dependent function is defined as ${\color{black}\eta^{\chi}_{\theta\phi}} = 2 \hbar v_{F} + 2 t^{\chi}_{x} \hbar v_{F} \sin{\theta} \cos{\phi} + 2 t^{\chi}_{z} \hbar v_{F} \cos{\theta}$, and the geometric factor is given by ${\color{black}\beta_{\theta \phi \gamma}} = \sin{\theta} \cos{\phi} \cos{\gamma} + \cos{\theta} \sin{\gamma}$. Interestingly, the angular profile of  $k^{\chi}_{F}$ indicates the emergence of non-spherical Fermi surface which is significantly different from a spherical Fermi surface of untilted WSM in the absence of OMM.
The direction of the magnetic field is measured with respect to the $\hat{x}$-axis, i.e., $\mathbf{B} = B (\cos\gamma, 0, \sin\gamma)$. The non-trivial topology of the system is encapsulated in the Bloch eigenstates of the Hamiltonian in Eq.~\eqref{Hamiltonian}, given by~\cite{knoll2020negative,xiao2010berry}: 
\begin{align}
\ket{u^{+1}_{\mathbf{k}}}^{T}&=[e^{-i\phi}\cos(\theta /2),~~\sin(\theta/2)],\nonumber \\
\ket{u^{-1}_{\mathbf{k}}}^{T}&=[-e^{-i\phi}\sin(\theta /2),~~\cos(\theta/2)].
\end{align}
These eigenstates generate a singular Berry curvature ${\mathbf \Omega}^{\chi}_{\mathbf k}=-\chi {\mathbf k}/(2k^3)$ and OMM ${\mathbf m}^{\chi}_{\mathbf k} = -{\chi e v_F\mathbf{k}}/{2k^2}$~\cite{xiao2010berry}. Furthermore, inclusion of OMM coupling to the external magnetic field modifies the band velocity {\color{black}$\mathbf{v}_\mathbf{k}^\chi = (1/\hbar)(\partial{\epsilon}^{\chi}_{\mathbf{k}}/\partial\mathbf{k})$}, and is expressed as:
\begin{align} 
v^{\chi}_{x,\textbf{k}} &= v_{F}\sin\theta\cos\phi + v_F t^{\chi}_x \nonumber \\
&+ \frac{C^{\chi}}{k^2} \left[\cos{\gamma}\left(1 - 2\sin^2\theta\cos^2\phi\right) -\sin{\gamma}\sin(2\theta)\cos\phi \right], \nonumber \\  
v^{\chi}_{y,\textbf{k}} &= v_{F}\sin\theta\sin\phi \nonumber \\
&- \frac{C^{\chi}}{k^2} \left[ \cos{\gamma}\sin^2\theta\sin(2\phi) + \sin{\gamma}\sin(2\theta)\sin\phi \right],\nonumber \\ 
v^{\chi}_{z,\textbf{k}} &= v_{F}\cos\theta + v_F t^{\chi}_z \nonumber \\
&- \frac{C^{\chi}}{k^2} \left[ \cos{\gamma}\sin(2\theta)\cos\phi + \sin{\gamma}\cos(2\theta) \right], \label{velocity_components_spherical} 
\end{align}
where $C^\chi={\chi e v_{F} B}/{2 \hbar}$ represents the field-dependent correction term.  
%
%
\section{Semiclassical Boltzmann Transport formalism} 
\label{sec:MBT}
%
%
To describe the dynamics of Weyl fermions in the presence of time-dependent electric and static external magnetic fields, we employ the semiclassical Boltzmann transport equation {\color{black}where the dynamics of Bloch electrons are governed by the following equations of motion~\cite{sundaram1999wave,son2012berry,knoll2020negative}:
\begin{align}
\dot{\mathbf{r}}_{\mathbf{k}}^\chi(t) &= \mathcal{D}^\chi_{\bf{k}} \left[ \frac{e}{\hbar}(\mathbf{E}(t)\times \boldsymbol{\Omega}_{\bf{k}}^\chi) + \frac{e}{\hbar}(\mathbf{v}_{\mathbf{k}}^\chi\cdot \boldsymbol{\Omega}_{\mathbf{k}}^\chi) \mathbf{B} + \mathbf{v}_\mathbf{k}^\chi\right], \nonumber\\
\dot{\mathbf{k}}_{\mathbf{k}}^\chi(t) &= -\frac{e}{\hbar} \mathcal{D}^\chi_{\bf{k}} \left[ \mathbf{E}(t) + \mathbf{v}_\mathbf{k}^\chi \times \mathbf{B} + \frac{e}{\hbar} (\mathbf{E}(t)\cdot\mathbf{B}) \boldsymbol{\Omega}_{\mathbf{k}}^\chi \right],
\label{Couplled_equation}
\end{align}
where, 
$\mathcal{D}^\chi_{\bf{k}} = (1+e\mathbf{B}\cdot\boldsymbol{\Omega}_\mathbf{k}^\chi/\hbar)^{-1}$ is a factor by which density of states is modified due to the presence of the Berry curvature~\cite{duval2006comment,xiao2010berry}.} In phase space, the non-equilibrium distribution function $f^{\chi}_{\mathbf{k}}$ evolves according to~\cite{kim2014boltzmann,knoll2020negative,imran2018berry,zyuzin2017magnetotransport}:
\begin{align}
\frac{\partial f^{\chi}_{\mathbf{k}}}{\partial t}
+ \Dot{\mathbf{r}}^{\,\chi}_{\mathbf{k}}(t) \cdot \boldsymbol{\nabla}_{\mathbf{r}} f^{\chi}_{\mathbf{k}}
+ \Dot{\mathbf{k}}^{\,\chi}_{\mathbf{k}}(t) \cdot \boldsymbol{\nabla}_{\mathbf{k}} f^{\chi}_{\mathbf{k}}
= I_{\text{coll}}[f^{\chi}_{\mathbf{k}}{\color{black}]}.
\label{MB_equation}
\end{align}
The distribution function is expanded as:~$f^{\chi}_{\mathbf{k}}(t)
= f^{\chi}_{0}(\epsilon_{\mathbf{k}}) 
+ g_{\mathbf{k}}^{\chi}(t)
+ \mathcal{O}(E^{2})$, with, $f^{\chi}_{0}$ is the equilibrium Fermi--Dirac distribution in the absence of any external field. We consider spatially uniform fields such that the distribution function $f^{\chi}_{\mathbf{k}}(t)$ only varies with time $t$ and momentum $\textbf{k}$.  
Here, the correction $g_{\mathbf{k}}^{\chi}(t)$ corresponds to the linear response to the time-dependent electric field {$\mathbf{E}(t)$}. 

For a time-dependent electric field, it takes the generic form~\cite{morimoto2016semiclassical,sodemann2015quantum,das2022nonlinear,mandal2022chiral}:
\begin{align}
g^{\chi}_{\mathbf{k}}(t) = -e \left[\mathbf{E}(t)\cdot\mathbf{\Lambda^{\chi}_k}\frac{\partial}{\partial {\color{black}\epsilon_{\bf{k}}^\chi}}\right] f_{0}^{\chi},
\label{Eq:g1}
\end{align}
with, $\mathbf{\Lambda}^{\chi}_{\mathbf{k}}$, having the dimension of length, is unknown function (ansatz) to be evaluated. For simplicity of the analytical calculations, \textcolor{black}{we first focus on the case of linearly polarized light}, where the electric field is given by $\mathbf{E}(t) = E e^{-i \omega t} ~\hat{e}_i + \textit{c.c.} $, $\hat{e}_i$ being the unit
vector along the $\hat{x}$-, or $\hat{y}$-, or $\hat{z}$-direction, $\omega$ is the drive frequency associated with the electromagnetic radiation. We illustrate the setup in Fig. \ref{fig:Exp_setup.png}. Under this condition, Eq.~\eqref{Eq:g1} reduces to \cite{das2022nonlinear,gao2022suppression}:
\begin{align}
g^{\chi}_{\bf k}(t) = -e \left( \frac{\partial f_{0}^{\chi}}{\partial {\color{black}\epsilon_{\bf{k}}^\chi}}\right) E_i(t)~\Lambda^{\chi}_{i, {\bf k} } .
\label{Eq:g1a}
\end{align}
We choose electric field along $\hat{z}$-direction i.e.,  $i=z$.    We consider that the collision integral term of the Eq.~\eqref{MB_equation} incorporates both internode ($\chi \neq \chi'$) and intranode ($\chi = \chi'$) electron-impurity scattering, and is expressed as~\cite{mahan20089,bruus2004many,ziman1979principles,knoll2020negative,ahmad2024geometry},
\begin{align}
 I_{\text{coll}}[f^{\chi}_{\mathbf{k}}]=\sum_{\chi' \mathbf{k}'}{\mathbf{W}^{\chi \chi'}_{\mathbf{k k'}}}{(f^{\chi'}_{\mathbf{k'}}-f^{\chi}_{\mathbf{k}})},
\label{Collision_integral}
\end{align}
where the scattering rate ${\mathbf{W}^{\chi \chi'}_{\mathbf{k k'}}}$ is calculated using Fermi's golden rule, \textcolor{black}{$\mathbf{W}^{\chi \chi'}_{\mathbf{k k'}} = \frac{2\pi n}{\hbar\mathcal{V}}|\bra{u^{\chi'}_{{\bf k}'}}U^{\chi \chi'}_{\mathbf{k k'}}\ket{u^{\chi}_\mathbf{k}}|^2 \times\delta(\epsilon^{\chi'}_{\mathbf{k'}}-\epsilon^{\chi}_{\mathbf{k}})$}~\cite{ziman1979principles,bruus2004many,mahan20089,abers2004quantum}. Here, \(n\) is the impurity concentration, \(\mathcal{V}\) is the system volume, \(U^{\chi\chi'}_{\mathbf{k}\mathbf{k}'}\) is the scattering potential profile. 
{ Given the structure of the Hamiltonian Eq. (\ref{Hamiltonian}), one can use the standard spherical polar co-ordinate to represent the above using the Pauli matrices $H= {\mathbf n}.{\mathbf \sigma}$ with ${\mathbf n}=(n_0,\sin\theta\cos\phi,\sin\theta\sin\phi,\cos\theta)$ and ${\mathbf \sigma}=(\sigma_0,\sigma_1,\sigma_2,\sigma_3)$. This representation enables one to write the wave functions in the form $\ket{u^{+1}_\mathbf{k}}=(\cos \theta/2, e^{i\phi}\sin \theta/2  )^T$ and $\ket{u^{-1}_\mathbf{k}}=(\sin \theta/2,- e^{i\phi} \cos \theta/2  )^T$, without loss of generality. }

In current formalism, even though our primary focus is on non-magnetic point-like impurities, one can choose \(U^{\chi\chi'}_{\mathbf{k}\mathbf{k}'}\) such that it can include both magnetic and non-magnetic point-like scattering centers~\cite{varma2024magnetotransport,varma2026chiral}. In general, \(U^{\chi\chi'}_{\mathbf{k}\mathbf{k}'} = U^{\chi\chi'} \sigma_i\) with \(i = 0, x, y, z\), where \(U^{\chi\chi'}\) distinguishes the internode and intranode scattering processes. Here, it is worth highlighting that in our formalism, the relative strength of these two scattering channels can be tuned using the dimensionless parameter defined as {\color{black}{$\alpha = \frac{U^{\chi\chi'}}{U^{\chi\chi}}$}}. Note that $g^{\chi}_{\bf k}(t)=g^{\chi}_{\bf k} e^{-i \omega t}$ where
$g^{\chi}_{\bf k}$ is the time-independent correction term derived from the amplitude of the ac electric field as given by 
{\begin{align}
g^{\chi}_{\bf k} = -e \left( \frac{\partial f_{0}^{\chi}}{\partial {\color{black}\epsilon_{\bf{k}}^\chi}}\right) E_z~\Lambda^{\chi}_{{z,\bf k} } .
\label{Eq:g1a1}
\end{align}}

To calculate $g^{\chi}_{\mathbf{k}}$, we can write the Boltzmann transport equation by using Eqs.~\eqref{Couplled_equation} and \eqref{Collision_integral}, and keeping terms up-to-linear order in the electric field as~\cite{bruus2004many,ziman1979principles,sharma2020sign,knoll2020negative,son2013chiral,sodemann2015quantum},
\begin{align}
&-i\omega g^{\chi}_{\mathbf{k}} -e{\color{black}\mathcal{D}^\chi_{\bf{k}}} \mathcal{A}^{\chi}_{\mathbf{k}}\frac{\partial f_0^\chi}{\partial \epsilon^\chi_\mathbf{k}}= \textcolor{black}{\sum\limits_{\chi',\mathbf{k}'} W^{\chi\chi'}_{\mathbf{k}\mathbf{k}'} (g^{\chi'}_{\mathbf{k}'} - g^{\chi}_\mathbf{k})},
\label{Eq_boltz21}
\end{align}
where $\mathcal{A^{\chi}_{\mathbf{k}}} = \left(\mathbf{E} + \frac{e\mathbf{E} \cdot\mathbf{B}}{\hbar} \boldsymbol{\Omega}_\mathbf{k}^\chi \right)\cdot \mathbf{v}^{\chi}_\mathbf{k}$. 
In low temperature limit, using $\sum_{\textbf{k}',\chi'} \to \mathcal{V}\int (\mathcal{D}^{\chi'}_{\textbf{k}})^{-1} d^3 k'/(2\pi)^3 \equiv \mathcal{V}/(2\pi)^3\iiint (\mathcal{D}^{\chi'}_{\textbf{k}'})^{-1} k'^2 \sin \theta' d\theta' d \phi' dk'  $ and exploiting the property of $\frac{\partial f_0^\chi}{\partial \epsilon^\chi_\mathbf{k}}$, Eq.~\eqref{Eq_boltz21} can be simplified over the Fermi surface {$\epsilon^{\chi}_{\mathbf{k}}=\epsilon^{\chi'}_{\mathbf{k}'}=\epsilon_F$} as,
\begin{align}
\label{Eq.:BTE_theta_phi}
 &  h^{\chi}_{z,\bf{k}}(\theta,\phi) + \frac{\Lambda^{\chi}_{z,\bf{k}}(\theta,\phi)}{\widetilde{\tau}^{\chi}_{\bf{k}}(\theta,\phi)} \\ \nonumber
 &=\sum_{\chi'} \Pi^{\chi\chi'}\iint \Bigg[\frac{(k'^{\chi'})^3 }{|\mathbf{v}^{\chi'}_{\mathbf{k}'}\cdot \mathbf{k}'^{\chi'}|} (\mathcal{D}^{\chi'}_{\mathbf{k}'})^{-1}  \Lambda_{z,\bf{k}'}^{\chi'}\Bigg]_{k'=k_F}   \mathcal{G}^{\chi \chi'}(\theta, \phi)  \nonumber \\
 & \times \sin\theta' d\theta' d\phi' \nonumber. 
\end{align}
Here, $\Pi^{\chi \chi'} = n/ 8\pi^3 \hbar^2$, $h^{\chi}_{z,\bf{k}}(\theta,\phi)=\mathcal{D}^{\chi}_{\mathbf{k}}[v^{\chi}_{z,\textbf{k}}+\frac{eB_z}{\hbar}(\mathbf{\Omega}^{\chi}_{k}\cdot \mathbf{v}^{\chi}_{\mathbf{k}})]$ with $B_z=B\sin{\gamma}$.  The overlap of the Bloch wave-function for non-magnetic point-like scattering centers, is given by the expression,
$\mathcal{G}^{\chi\chi'}(\theta,\phi) = \mathcal{V}^2|U^{\chi\chi'}|^2[1+\chi\chi'(\cos{\theta}\cos{\theta'} + \sin{\theta}\sin{\theta'}\cos{\phi}\cos{\phi'} + \sin{\theta}\sin{\theta'}\sin{\phi}\sin{\phi'})]$. {Given the spherical structure of the Fermi surface as introduced by $(\partial f_{0}^{\chi}/\partial\epsilon_{\bf{k}}^\chi)$, we can obtain the above expression of the overlap function $\mathcal{G}^{\chi\chi'}(\theta,\phi)$.
} Furthermore, the inverse of the dressed relaxation time under the ac electric field is given by $1/\widetilde{\tau}^{\chi}_{\bf{k}}(\theta,\phi) = \bigg(\frac{1}{\tau^{\chi}_{\bf{k}}(\theta,\phi)} -i\omega \bigg)$ with 
\begin{align}
 \frac{1}{\tau^{\chi}_{\mathbf{k}}(\theta,\phi)}&= \sum_{\chi'} \Pi^{\chi\chi'}\iint \Bigg[\frac{(k'^{\chi'})^3}{|\mathbf{v}^{\chi'}_{{\bf k}'}\cdot{\mathbf{k}^{\chi'}}|} (\mathcal{D}^{\chi'}_{\mathbf{k}'})^{-1} \Bigg]_{k'=k_F}  \nonumber \\ & \times \mathcal{G}^{\chi\chi'}(\theta, \phi) \sin{\theta'}d\theta'd\phi',
\label{Tau_inv_int_thet_phi}
\end{align}
where the relaxation time is evaluated on the Fermi surface and varies with $\theta$ and $\phi$. This signifies the role of deformed-spherical Fermi surface in determining  the effects of tilts and OMM on the transport coefficients.  It is important to note that the relaxation time  in the absence of OMM for untilted WSM depends only on $\theta$ due to the perfectly spherical nature of the Fermi surface. 
%
\begin{figure}
    \centering
  \includegraphics[width=.95\columnwidth]{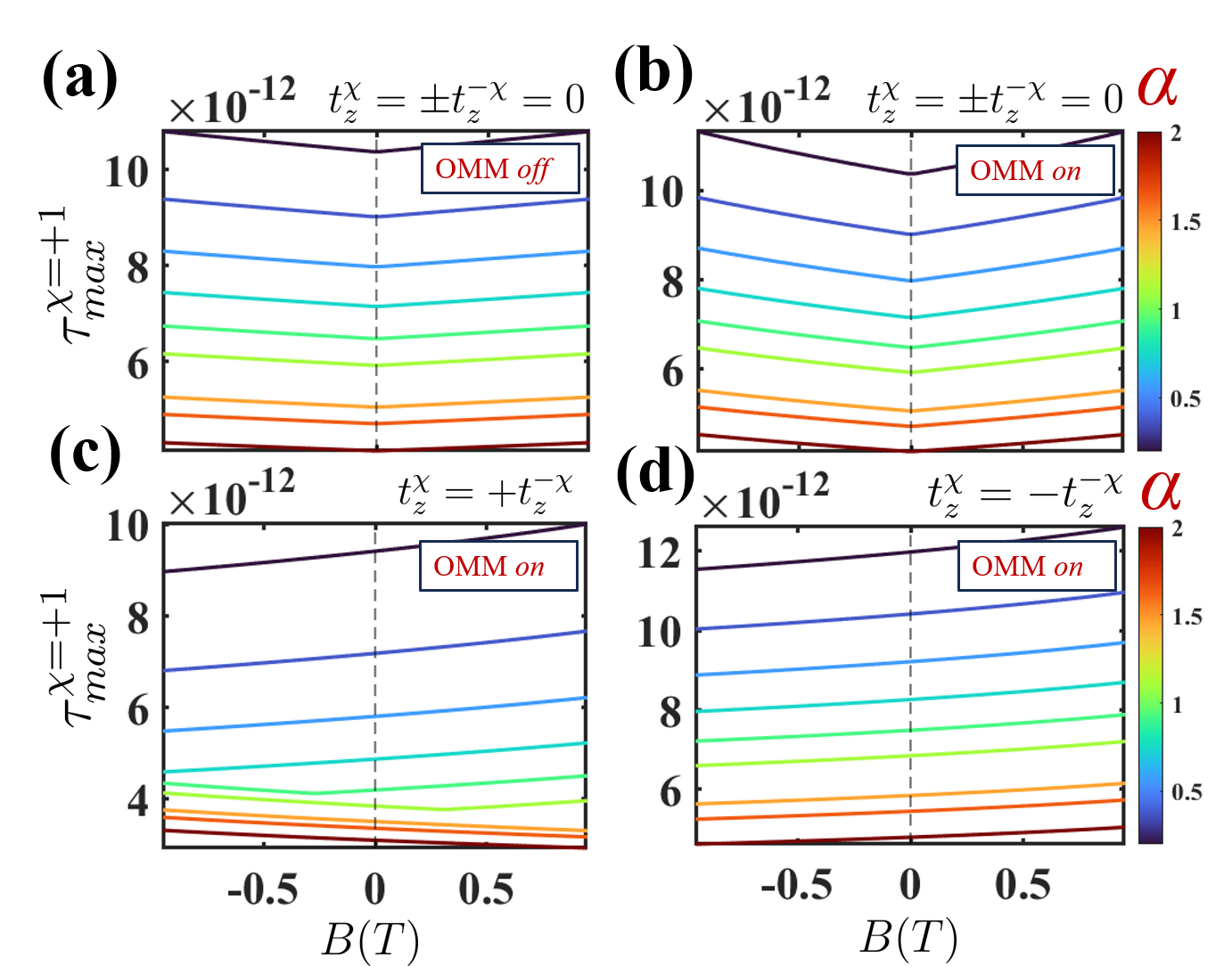}
    \caption{We show the variation of characteristic time scale $\tau$, associated with chirality $\chi=+$ as a function of $ B$ under OMM off and on conditions in (a) and (b), respectively, for untilted Weyl cones with $t_z^{\chi}= \pm t_z^{-\chi}=0$.  This is repeated for identically tilted $t_z^{\chi}=  t_z^{-\chi}\ne 0$ and oppositely tilted $t_z^{\chi}= - t_z^{-\chi}\ne 0$ Weyl cones in  (c), and (d), respectively,  in the presence of OMM.  
    Here, max in the subscript indicates  the highest value among all $\tau^{+}_{\mu}(\theta,\phi)$, i.e., $\tau^{+}_{\rm max} = {\rm max}\{\tau^{+}_{\mu}(\theta,\phi)\}$. For tilted case, tilt magnitude is $t_z^{\chi} = \pm t_z^{-\chi} = 0.45v_F$.}
\label{fig:tau1_vs_B_alp_vary.png}
\end{figure}
%

Motivated by the angular dependencies of the overlap function  $\mathcal{G}^{\chi\chi'}(\theta,\phi)$,  one can choose the following ansatz $\Lambda^{\chi}_{z,\mathbf{k}}=\big[d^{\chi}-h^{\chi}_{z,\mathbf{k}} + a^{\chi}\cos{\theta} +b^{\chi}\sin{\theta}\cos{\phi}+c^{\chi}\sin{\theta}\sin{\phi} \big] \textcolor{black}{\widetilde{\tau}^{\chi}_{\textbf{k}}(\theta,\phi)}$, where $a^\chi$, $b^\chi$, $c^{\chi}$ and  $d^\chi$ are the unknown coefficients having the dimension of velocity.
{We note that the Fermi surface of Weyl semimetals has in general a spherical symmetry. 
The key point is that the unknown quantity $\Lambda^{\chi}_{z,k}$ depends only on the spherical coordinates $(\theta,\phi)$ on the Fermi surface. The collision kernel entering Eq. (\ref{Eq.:BTE_theta_phi}), including the overlap factor $G^{\chi\chi'}(\theta',\phi',\theta,\phi)$, contains only the angular structures $[ 1,\cos\theta,\sin\theta\cos\phi,\sin\theta\sin\phi]$, which correspond to the ($l=0$) and ($l=1$) spherical-harmonic sector. Therefore, represents the most general expansion of $(\Lambda^{\chi}_{z,k})$ within the closed angular sector relevant to the problem obeying the symmetry and harmonic structure of the Boltzmann kernel. }
The  above Eq. (\ref{Eq.:BTE_theta_phi}) can be written in following form
{\color{black}
\begin{align}
&d^{\chi}+a^{\chi}\cos{\theta}+b^{\chi}\sin{\theta}\cos{\phi}+c^{\chi} \sin{\theta}\sin{\phi} \nonumber \\
&=\sum_{\chi'}\Pi^{\chi\chi'}\iint \Bigg[\frac{(k'^{\chi'})^3}{|\mathbf{v}^{\chi'}_{{\bf k}'}\cdot{\mathbf{k}^{\chi'}}|} \widetilde{\tau}^{\chi'}_{\textbf{k}'}(\theta,\phi) (\mathcal{D}^{\chi'}_{\mathbf{k}'})^{-1}\Bigg]_{k'=k_F}   \nonumber \\
&\times\bigg[ d^{\chi'}-\big[h^{\chi'}_{z,\textbf{k}'}\big]_{k'=k_F}+a^{\chi'}\cos{\theta'}+b^{\chi'}\sin{\theta'}\cos{\phi'}  \nonumber \\ 
&+c^{\chi'} \sin{\theta'}\sin{\phi'} \bigg]
 \mathcal{G}^{\chi\chi'} (\theta, \phi) ~\sin{\theta'}d\theta'd\phi'.
\label{Boltzman_final}
\end{align}
}
After equating the coefficients associated with $(1, \cos\theta, \sin\theta \cos \phi, \sin\theta \sin \phi)$ terms from both sides of the above Eq. (\ref{Boltzman_final}), one can solve the coefficients using a matrix form. 
Once this set of equations is expanded for $\chi=\pm1$, it actually yields seven linearly independent equations consisitng eight variables that must be solved.  The particle number conservation provides an additional restriction. The particle number conservation is given by $\sum\limits_{\chi}\sum\limits_{\mathbf{k}} g^{\chi}_\mathbf{k}= 0$ which takes the following form 
\begin{align}
&\sum_{\chi}\iint  \Bigg[\frac{(k^\chi)^3\widetilde{\tau}^{\chi}_{\bf{k}}(\theta,\phi)}{|\mathbf{v}^{\chi}_{{\bf k}}\cdot{\mathbf{k}^{\chi}}|} \Bigg]_{k=k_F} \sin{\theta} ~ 
\bigg[d^{\chi}-\big[h^{\chi}_{z,\textbf{k}}\big]_{k=k_F}\nonumber \\
&
+a^{\chi}\cos{\theta}+b^{\chi}\sin{\theta}\cos{\phi}+c^{\chi} \sin{\theta}\sin{\phi}\bigg]~ d\theta d\phi= 0.
\label{Eq_sumgk}
\end{align}
For eight unknown coefficients ($d^{\pm}, a^{\pm }, b^{\pm }, c^{\pm }$), Eqs. (\ref{Boltzman_final}) and (\ref{Eq_sumgk}) are simultaneously solved with Eq.~(\ref{Tau_inv_int_thet_phi}). Due to the intricate structure of equations, all two-dimensional integrals with respect to $\theta'$ and $\phi'$, the solutions of the coupled equations are carried out numerically. {Here we would like to emphasize that as a validation of the formalism, we show in Appendix-\ref{appB} that the conductivity expressions reduce to the known analytical results in the $B =0$ limit. In addition, the weak-ac limit continuously reproduces the previously reported dc magnetotransport behavior. The numerical calculations were checked for convergence by increasing the angular discretization used in the Fermi-surface integrations; the conductivity values and all qualitative features reported in this work were found to be stable against further refinement.}

Having discussed the linearly polarized light extensively, we now briefly discuss the case of  circularly polarized light   where ac electric field is given by $\mathbf{E}(t) = E e^{-i \omega t} (\hat{e}_l +  i \eta \hat{e}_m)/\sqrt{2} + \textit{c.c.} $, and the helicity $\eta=\pm$ denotes the right and left circular polarization with $l\ne m$. Here $l=z,m=x$ represent two orthogonal directions. This yields the correction term  
\begin{align}
 g^{\chi}_{\bf k} &= -e \left( \frac{\partial f_{0}^{\chi}}{\partial \epsilon_{\bf{k}}^\chi}\right) [E_z~\Lambda^{\chi}_{z, {\bf k}} \pm i E_x~\Lambda^{\chi}_{x, {\bf k} } ] \nonumber \\
&= g^{\chi}_{z,\bf k} \pm i g^{\chi}_{x,\bf k},
\label{Eq:g1a_cpl}
\end{align}
while considering the ansatz $\Lambda^{\chi}_{i,\mathbf{k}}=\big[d^{\chi}_i-h^{\chi}_{i,\mathbf{k}} + a^{\chi}_i\cos{\theta} +b^{\chi}_i\sin{\theta}\cos{\phi}+c^{\chi}_i\sin{\theta}\sin{\phi} \big] \textcolor{black}{\widetilde{\tau}^{\chi}_{\textbf{k}}(\theta,\phi)}$
with $i=z,x$. We have to solve two-component $g^{\chi}_{z/x,\bf k}(t)$ using two sets of  Eq. (\ref{Eq_boltz21}). One obtains sixteen unknown coefficients ($d^{\pm}_i, a^{\pm }_i, b^{\pm }_i, c^{\pm }_i$) with $i=z,x$  in the case of circularly polarized light.  See Appendix  \ref{sec:Generalization to circularly polarised light} for more details.

After evaluating the distribution function, the current density can be obtained upto the linear order in electric field as: $\mathbf{J}(t) =-e\sum_{\chi,\mathbf{k}} f^{\chi}_{\mathbf{k}}(t) ~\dot{\mathbf{r}}_{\mathbf{k}}^{\chi}(t)/ \mathcal{D}^\chi_{\bf{k}}= \sigma\textbf{E}(t)$ where $f^{\chi}_{\mathbf{k}}(t)= f^{\chi}_{0}(\epsilon_{\mathbf{k}}) 
+ g_{\mathbf{k}}^{\chi}(t)$. Note that the correction term  $g^{\chi}_{\mathbf{k}}$ is a scalar quantity and chosen from Eqs. (\ref{Eq:g1a1}) and (\ref{Eq:g1a_cpl}) for linearly  and circularly polarized light, respectively.  The transverse and longitudinal  components of conductivities  $\sigma_{\alpha \beta}$ are expressed below with $\alpha\ne \beta, \alpha=x,\beta=z$,  and $\alpha=\beta=z$, respectively.

The general expression of linear magneto-optical conductivity  can be found as 
\begin{align}
    \sigma_{xz} &=\frac{-e^2}{\hbar} \sum_{\chi}\int\frac{d^3\mathbf{k}}{(2\pi)^3} \bigg[ \left(  ( \mathbf{v}_\mathbf{k}^\chi \cdot \boldsymbol{\Omega}^\chi_{\mathbf{k}})~ B \cos\gamma +  \frac{\hbar}{e} v_{x,\textbf{k}}^{\chi}\right)  \nonumber \\
    & \times \Lambda^{\chi}_{z, {\bf k}}\left( \frac{\partial f_{0}^{\chi}}{\partial \epsilon_{\bf{k}}^\chi}\right) + \Omega_{y,\textbf{k}}^\chi ~f^\chi_{0}  \bigg],\nonumber\\ 
    \sigma_{zz}  &=\frac{-e^2}{\hbar} \sum_{\chi}\int\frac{d^3\mathbf{k}}{(2\pi)^3} \bigg[ \left(  ( \mathbf{v}_\mathbf{k}^\chi \cdot \boldsymbol{\Omega}^\chi_{\mathbf{k}})~ B \sin\gamma + \frac{\hbar}{e} v_{z,\textbf{k}}^{\chi}\right)  \nonumber \\ &\times ~\Lambda^{\chi}_{z, {\bf k}}\left( \frac{\partial f_{0}^{\chi}}{\partial \epsilon_{\bf{k}}^\chi}\right) \bigg].
    \label{Eq:current linear response}
\end{align}
Importantly, in the present case of an ac electric field, associated with the linearly polarized light $\Lambda^{\chi}_{z, {\bf k}}(\omega)$ depends on the frequency of the electric field. We, therefore, obtain  LMOC $\sigma_{\alpha \alpha}(\omega)$ and TMOC $\sigma_{\alpha \beta}(\omega)$ that depends on frequency. 
For ease of notation, we will henceforth refer to $\sigma_{\alpha\beta}(\omega)$  as $\sigma_{\alpha\beta}$. For completeness, we derive the magneto-optical conductivities in  the absence of an external magnetic field $B=0$ that match with the literature \cite{sodemann2015quantum,zhang2023higher,gao2022suppression,gupta2024magneto,dagnino2025non,prabhat2026magneto}, see  Appendix \ref{appB} for more details.

It is important to note that we consider incident electromagnetic radiation in the frequency window where the response of Weyl fermions is well described by semiclassical dynamics, namely $\omega \ll \epsilon_F/\hbar$~\cite{heidari2020chiral}. Within this regime, we systematically vary the driving frequency to analyze its influence on the magneto-optical response without perturbing the underlying Hamiltonian. The relevant transport regimes are classified according to the dimensionless parameter $\omega \tau$, where $\tau\equiv \tau^\chi={\rm max}\{\tau^\chi_{\textbf{k}}(\theta,\phi)\}$ denotes the largest relaxation time over the $(\theta,\phi)$-plane in the system and is plotted in Fig.~\ref{fig:tau1_vs_B_alp_vary.png} as a function of $B$ with different values of $\alpha$ and tilt strength. We focus on three qualitatively distinct regimes depending on the frequency of the ac electric field: (i)  weak ac limit with low frequency such that  $\omega\tau \ll 1$, (ii)  moderate ac limit with intermediate frequency such that $\omega\tau \sim 1$, and (iii) strong ac limit with high frequency such that $\omega\tau > 1$. {Note that, to classify the frequency response, we use the largest relaxation time on the Fermi surface, $\tau^\chi={\rm max}\{\tau^\chi_{\textbf{k}}(\theta,\phi)\}$, as a convenient reference scale. The weak-, intermediate-, and strong-ac regimes correspond approximately to $\omega \tau^\chi \ll 1$ , $\omega \tau^\chi \sim 1$, and $\omega \tau^\chi \geq 1$, respectively. The crossover is gradual, and alternative definitions based on averaged relaxation times would lead to quantitatively similar classifications. One can consider $\tau^\chi$ from the average or minimal relaxation time on the Fermi surface such as,
$\tau^\chi={\rm avg}\{\tau^\chi_{\textbf{k}}(\theta,\phi)\}$ or $\tau^\chi={\rm min}\{\tau^\chi_{\textbf{k}}(\theta,\phi)\}$.  However, the results do not alter qualitatively as long as the conductivities are computed deep inside the above regimes. } 

\begin{figure*}
    \centering
    \includegraphics[width=1.9\columnwidth]{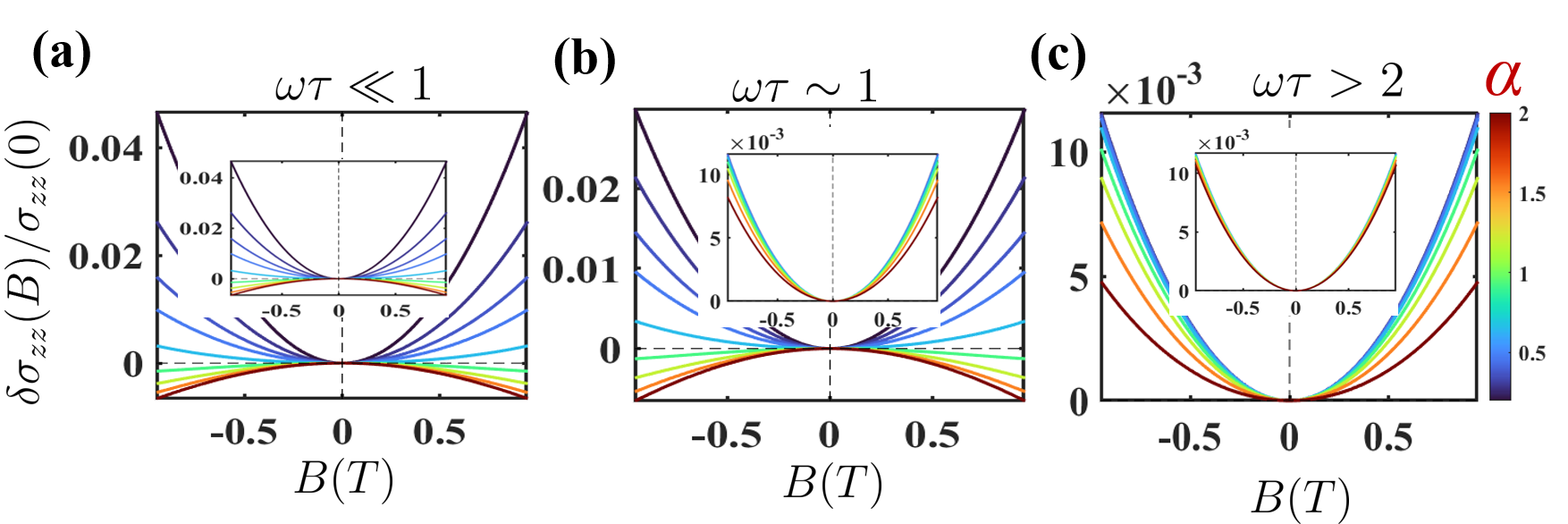}
    \caption{LMOC namely, the real part of the quantity $\delta \sigma_{zz}(B)/\sigma_{zz}(0)={\rm Re}[\sigma_{zz}(B\ne 0)/\sigma_{zz}(0)-1]$, of a untilted WSM in the presence of OMM with non-magnetic impurity $U^{\chi\chi'}_{\mathbf{k}\mathbf{k}} = U^{\chi\chi'} \sigma_0$ under the application of linearly polarized light. (a) In the weak ac limit ($\omega\tau \ll 1$), a sign reversal with increasing intervalley scattering strength is observed, consistent with the suppression of the anomaly-induced contribution by strong intervalley relaxation~\cite{ahmad2021longitudinal,ahmad2025longitudinal,sharma2020sign}. (b) illustrates the moderate ac limit where LMOC reverses its sign for large intervalley scattering similar to that of (a). (c) does not show any sign reversal of LMOC  for large  $\alpha$ in the strong ac limit, indicating the dynamical suppression of intervalley scattering. 
  The inset shows the behavior of LMOC, obtained using under magnetic impurity scattering, $U^{\chi\chi'}_{\mathbf{k}\mathbf{k}} = U^{\chi\chi'} \sigma_x$,  in the respective frequency window. The magnetic field is applied at $\gamma=\pi/2$ with respect to the $\hat{x}$ direction. {Note that we consider $100\times 150$ grid for $\theta\in[0, \pi]$ and $\phi \in [0, 2\pi]$ for all our numerical calculations.} }
    \label{LMC_vs_B_apl_vary_omg_vary.fig.png}
\end{figure*}

One can distinguish the driving frequency $\omega$ from the cyclotron frequency associated with an external magnetic field. The cyclotron frequency, associated with the static magnetic field,  is given by $\omega_c = eB/m^*$, where the cyclotron effective (dynamical) mass of a Weyl fermion is $m^* = \epsilon_F/v_F^2$~\cite{roy2016universal}. 
In the perturbative regime considered here, the magnetic component of the electromagnetic radiation is very small, $B^{\mathrm{EM}} \sim 10^{-6}\,\mathrm{T}$, as discussed earlier. This field is too weak to induce
any appreciable modification of the semiclassical phase-space factor
$\mathcal{D}^\chi_{\bf{k}} = (1+e\mathbf{B^{\mathrm{EM}}}\cdot\boldsymbol{\Omega}_\mathbf{k}^\chi/\hbar)^{-1}$
or to generate a measurable magnetic-field-induced contribution to the conductivities. 
Note that the associated cyclotron frequency $\omega_c^{\mathrm{EM}} = e B^{\mathrm{EM}}/m^* \sim 10^{7}\,\mathrm{Hz}$ is significantly smaller than $\omega_c$. Most importantly, $ \omega_c \ll \epsilon_F/\hbar$, ensuring that Landau quantization due to the static field is negligible and the semiclassical description remains valid. Unless otherwise stated, we use the following parameter values: Fermi velocity $v_F = 10^{6}\,\mathrm{m/s}$, Fermi energy $\epsilon_F = 0.05\,\mathrm{eV}$, reduced Planck constant $\hbar = 1.0\times10^{-34}\,\mathrm{J\,s}$, and electronic charge $e = 1.6\times10^{-19}\,\mathrm{C}$. With these choices, the hierarchy of energy and frequency scales employed in our analysis is internally consistent.
\section{Results and discussions} 
\label{r_results}

In this section, we numerically evaluate the longitudinal magneto-conductivity and planar Hall conductivity by performing integrations over the Berry curvature–modified
phase space. In the absence of an external magnetic field, the phase-space
correction factor reduces to unity, and the longitudinal conductivity is
entirely governed by Fermi-surface contributions, recovering the conventional Drude form
for $\sigma_{zz}$. In this limit, Berry curvature does not contribute to the longitudinal response. In the presence of a finite static magnetic field with $\mathbf{E}\parallel\mathbf{B}$, the longitudinal conductivity remains positive and exhibits a clear enhancement with a characteristic dependence on $B^{2}$, serving as a hallmark signature of the chiral anomaly in WSMs. To focus on the study of the effect of electromagnetic  radiation on WSMs in the presence of an external magnetic field, we ignore the anomalous contribution to the conductivity in the present analysis.

Note that dressed relaxation time $\widetilde{\tau}$ contains an imaginary contribution, the conductivities have both real and imaginary components.   The imaginary nature of the off-diagonal component of the conductivity  is primarily reactive at finite frequency. Importantly,  the physically measurable LMOC is determined by the real i.e., dissipative part of the conductivity. 
We examine the real part of the quantity $\delta \sigma_{\alpha\beta}(B)/\sigma_{\alpha\beta}(0)={\rm Re}[\sigma_{\alpha\beta}(B\ne 0)/\sigma_{\alpha\beta}(0)-1]$ by discarding the Drude contribution in all our future investigations on LMOC. 
Furthermore, since the tilt qualitatively modifies the transport coefficients, we study the tilted and untilted cases separately to distinguish their effects on the conductivity in Secs. \ref{res1} and \ref{res2}, respectively. It is noteworthy that our formalism can accurately reproduce the conductivities  at zero magnetic field, see Appendix \ref{appB} for more details.

\subsection{Magneto-optical response for untilted Weyl cones}
\label{res1}


\begin{figure}
    \centering
    \includegraphics[width=.95\columnwidth]{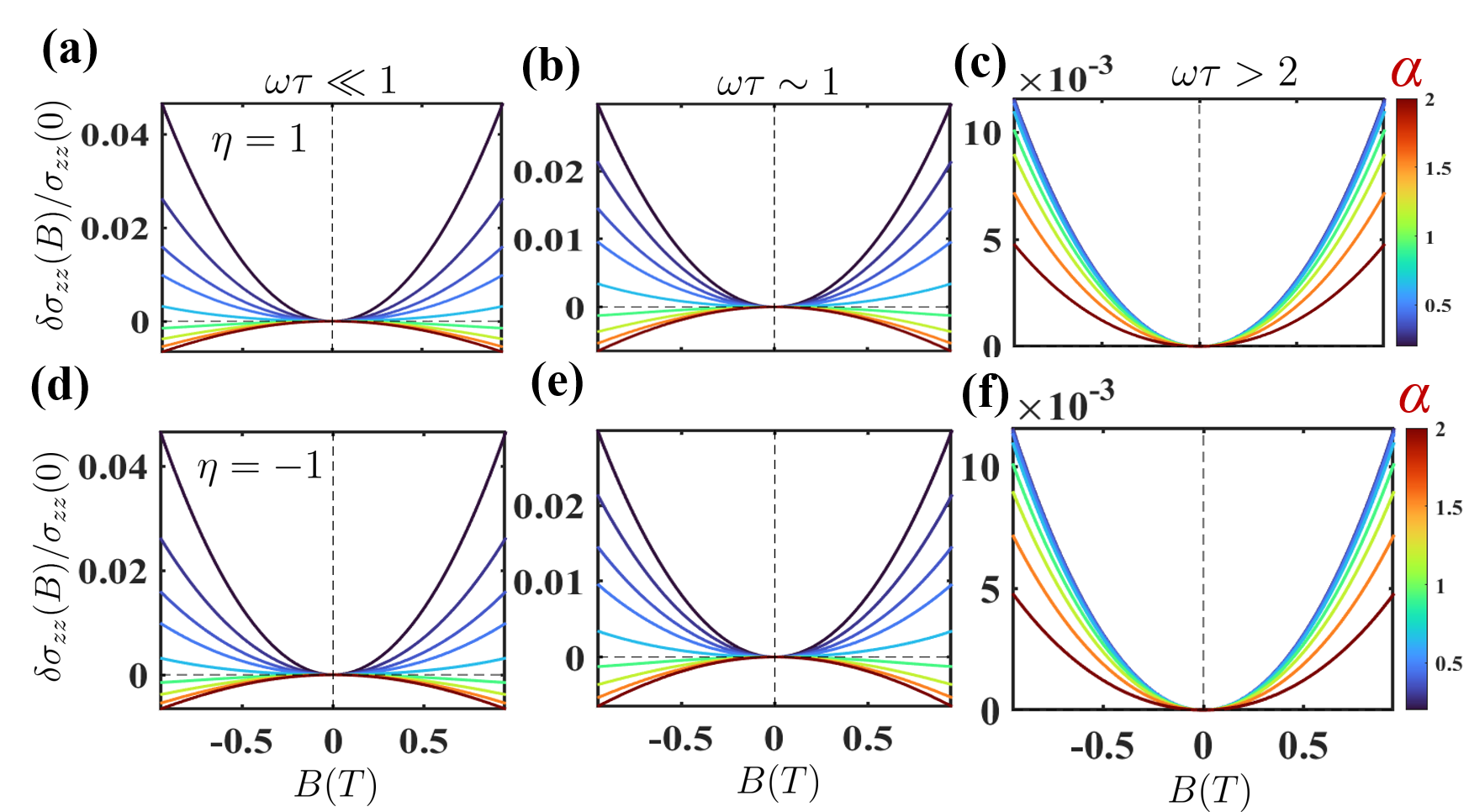}
    \caption{We repeat Fig. \ref{LMC_vs_B_apl_vary_omg_vary.fig.png} under the application of circularly polarized light. We show the LMOC for right and left circular polarization with $\eta=+1$ and $-1$ in (a,b,c) and (d,e,f), respectively. Similar to the linear polarization, a sign reversal occurs with increasing intervalley scattering strength only in the weak and intermediate regimes while LMOC is thoroughly positive in the strong ac regime. }
    \label{fig:LMC_vs_B_CPEG.png}
\end{figure} 
In this section, we consider the untilted WSM described by the Hamiltonian in Eq. (\ref{Hamiltonian})  with $t^{\chi}_{x,z}=0$ for the analysis of magneto-optical conductivities. 
The longitudinal current reveals that $J_z$ is governed by two distinct mechanisms. The first is the conventional Drude contribution associated with field-driven carrier transport near the Fermi surface,  whereas the second one, vanishing at zero magnetic field, originates from Berry curvature–induced modifications. The Berry curvature appears in the semiclassical equations of motion and the phase-space measure through the $(\mathbf{v}\!\cdot\!\boldsymbol{\Omega})\mathbf{B}$ and the phase-space correction  $\mathcal {D}$ terms, respectively, giving rise to an anomaly-related contribution proportional to $(\mathbf{E} \cdot \mathbf{B})\mathbf{B}$~\cite{gao2022suppression,gupta2024magneto}.    Figure~\ref{LMC_vs_B_apl_vary_omg_vary.fig.png} displays the LMOC of a WSM driven by a monochromatic electric field in the presence of a static magnetic field applied at an angle $\gamma=\pi/2$ with respect to $\hat{x}$. Considering non-magnetic impurity, the LMOC in the presence of OMM is shown as a function of {dimensionless parameter} $\alpha=\frac{U^{\chi\chi'}}{U^{\chi\chi}}$,  for weak, moderate and strong ac limits in Figs. \ref{LMC_vs_B_apl_vary_omg_vary.fig.png} (a,b,c), respectively.

In the weak ac regime ($\omega\tau \ll 1$), 
the LMOC contains both a positive chiral anomaly–induced contribution and a conventional Drude term. In the presence of OMM, the competition between the anomaly-driven current and the relaxation of chiral charge via intervalley scattering leads to a sign reversal of the LMOC beyond a critical scattering strength $\alpha_c$ as shown in panel ~\ref{LMC_vs_B_apl_vary_omg_vary.fig.png}(a).  This behavior is consistent with the known dc response and serves as a benchmark for our finite-frequency analysis~\cite{ahmad2021longitudinal,ahmad2025longitudinal,sharma2020sign}.

In the intermediate dynamical regime $\omega\tau\sim1$, as shown in Fig. \ref{LMC_vs_B_apl_vary_omg_vary.fig.png}(b), the driving frequency becomes comparable to the relaxation rate. Here, the nonequilibrium distribution function acquires a frequency dependence through the effective relaxation time $\widetilde{\tau}_c(\omega)=(\tau_c^{-1}-i\omega)^{-1}$. In this regime, the degree of chiral imbalance is partially suppressed which is  evident by the reduction in magnitude of the response $\delta \sigma_{\alpha\beta}(B)/\sigma_{\alpha\beta}(0)$
as compared to the weak ac limit. Otherwise, the responses for a non-magnetic impurity in the weak and moderate ac regimes are qualitatively identical. 
In Fig. \ref{LMC_vs_B_apl_vary_omg_vary.fig.png}(c), corresponding to the strong ac regime $\omega\tau>2$, we find that the driving period of the electric field is much shorter than the relevant scattering times for the intervalley processes.  Therefore, the chiral imbalance within a single optical cycle cannot relax with the time scale, rendering the LMOC to remain positive under the variation of intervalley scattering strength. In this limit, the positive  longitudinal response is governed by intrinsic semiclassical dynamics, exhibiting $B^{2}$ scaling. Interestingly, OMM and substantially strong intervalley scattering are unable to reverse the sign of the LMOC in the strong ac limit, which is markedly distinct compared to the dc response of longitudinal magnetoconductivity.

Further, the inclusion of magnetic impurity scattering, characterized by $U^{\chi\chi'}_{\mathbf{k}\mathbf{k}} = U^{\chi\chi'} \sigma_x$, does not qualitatively modify the LMOC, see the insets in Figs. \ref{LMC_vs_B_apl_vary_omg_vary.fig.png}(a,b,c) for weak, moderate and strong ac limit, respectively. Surprisingly,  the sign reversal phenomenon of LMOC ceases early in terms of the frequency of the ac electric field for the magnetic impurity as compared to the non-magnetic one, see the Fig. \ref{LMC_vs_B_apl_vary_omg_vary.fig.png}(b) and its inset, where there is no sign reversal occurring at moderate frequency. In the strong limit, the LMOC remains completely insensitive to $\alpha$ for the magnetic impurity which is also expected to appear for the non-magnetic case, but for much larger frequency. Therefore, the responses with frequency changes quantitatively while keeping their qualitative behavior unchanged between magnetic and non-magnetic impurities. For completeness, we examine the imaginary part of the response in  Fig. \ref{fig:Imag_LMC_vs_B_alp_vary1} of Appendix \ref{appc}.

We now analyze the LMOC for circularly polarized light   within the same semiclassical Boltzmann framework. As expected, the $\sigma_{zz}$ and $\sigma_{zx}$ both contribute to the current $J_z$ due to the presence of $E_z$ and $E_x$ fields, respectively.  Notice that  the anomaly-induced contribution originates from the $(\mathbf{E}\!\cdot\!\mathbf{B})\mathbf{B}$ term, which depends only on the projection $E_z B_z$. Interestingly, 
$\delta \sigma_{zz}(B)/\sigma_{zz}(0)$ is independent of the helicity $\eta$ of the electromagnetic radiation resulting in an identical response of LMOC under both types of polarizations, see Figs.~\ref{fig:LMC_vs_B_CPEG.png} (a,b,c), and (d,e,f),  respectively, for right and left circularly polarized light. We find an identical profile of LMOC for right and left circularly polarized light in weak, moderate and strong ac limits, see  Figs.~\ref{fig:LMC_vs_B_CPEG.png} (a,d), (b,e), and (c,f), respectively.
 The LMOC for circularly polarized light behaves identically to that of linearly polarized light as $\sigma_{zz}$ expression remains unaltered irrespective of the choice of polarization. It is noteworthy that the helicity-dependent transport is naturally expected in the nonlinear regime ~\cite{sodemann2015quantum}, whereas the linear responses are helicity-independent. For completeness, we examine the imaginary part of LMOC for right and left circularly polarized light in  Fig. \ref{fig:Imag_LMC_vs_B_alp_vary1_eta_pm1} of Appendix \ref{appc}.

\begin{figure}
    \centering
    \includegraphics[width=.95\columnwidth]{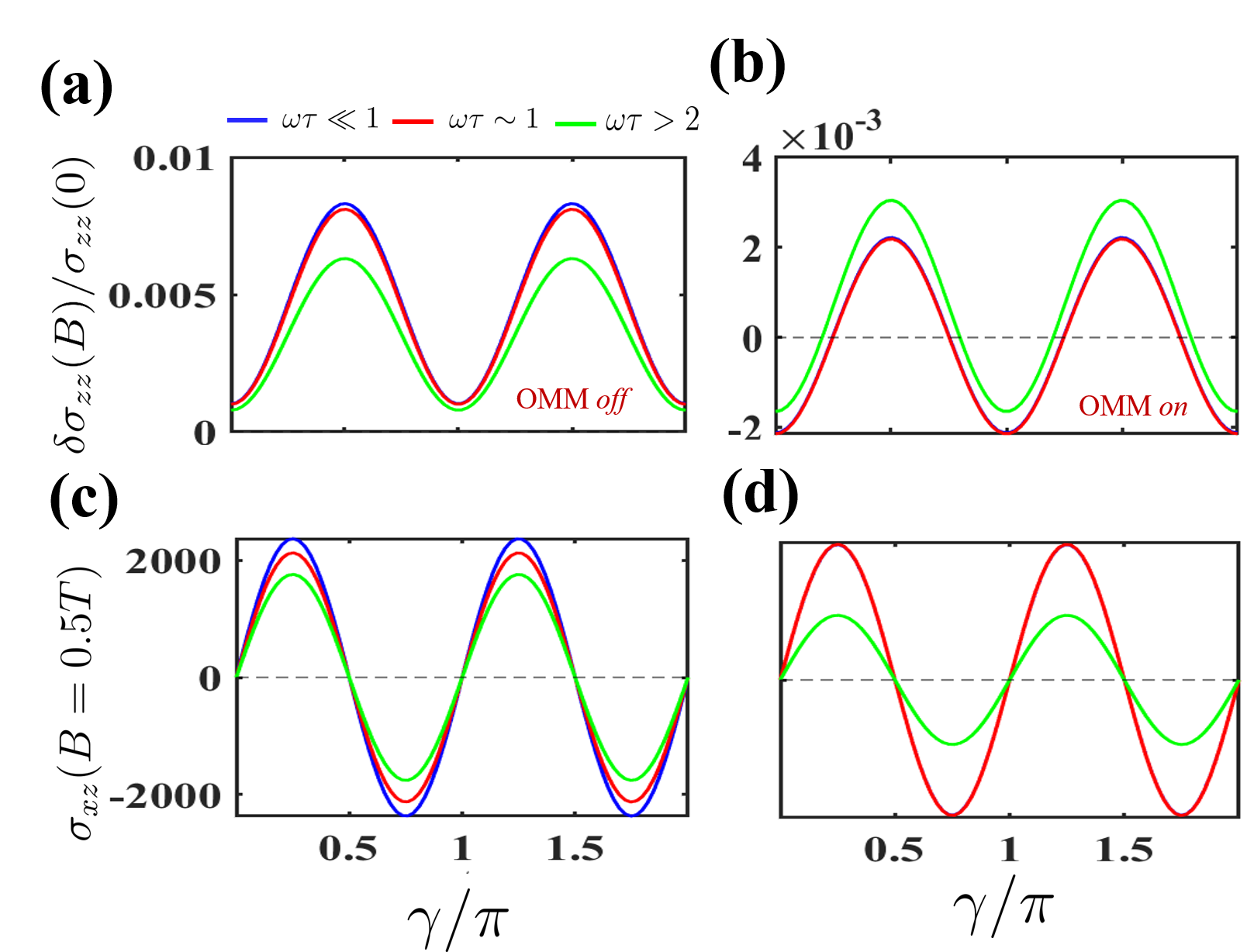}
    \caption{We show the angular variation of LMOC and TMOC in (a,b) and (c,d), respectively, under the application of linearly polarized light in the presence of OMM. We illustrate the response in three different regimes $\omega \tau \ll1$, $\sim 1$, and $> 2$ with blue, red, and green lines, respectively. The static magnetic field $(B\cos \gamma, 0, B \sin\gamma)$ and electric field $(0,0,E_z)$, associated with the electromagnetic radiation, lie on the $xz$-plane causing the 
    $\sin^2\gamma$ and $\sin 2\gamma$ response for LMOC and TMOC, respectively. In (c) and (d), the TMOC under the moderate and strong ac regimes, depicted by red and green lines,  has been scaled by 5 times and 30 times, respectively. In all the plots, we consider $\alpha=0.50$.}
    \label{fig:LMC_PHC_vs_gm.png}
\end{figure}


\begin{figure*}[ht]
    \centering
    \includegraphics[width=1.95\columnwidth]{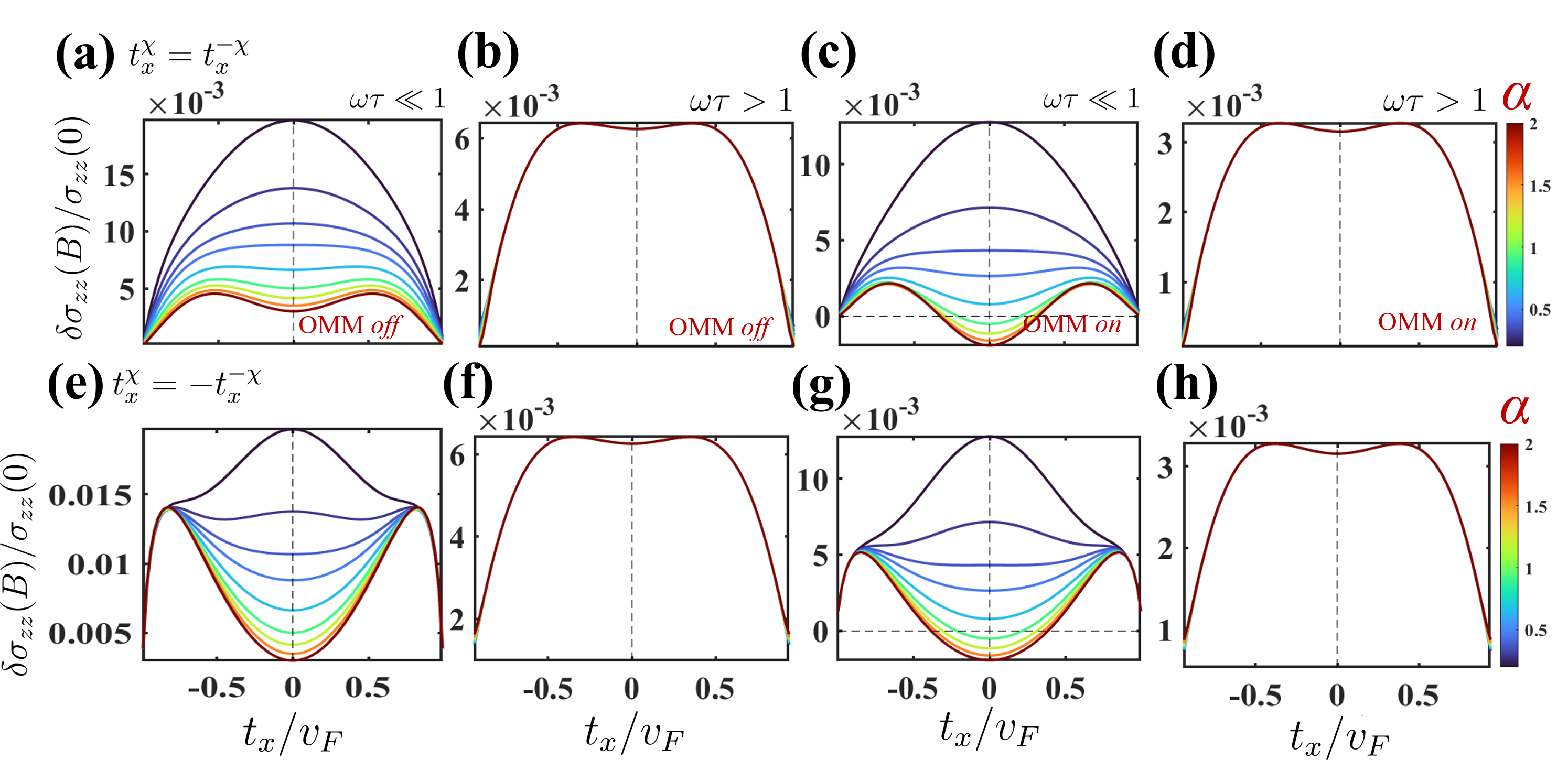}
    \caption{We show the variation of LMOC for the tilt direction perpendicular to the applied magnetic field i.e., $t_x^{\chi} \perp B_z$ under linearly polarized electromagnetic radiation with $E_z$. We examine two situations with identical and opposite tilt orientations of the Weyl cones with $t_x^{\chi}= t_x^{-\chi}$ and $t_x^{\chi}= -t_x^{-\chi}$ in (a,b,c,d) and (e,f,g,h), respectively. We analyze the weak and strong ac limit, both in the presence and absence of OMM which causes the sign reversal of LMOC with strong intervalley scattering.  
    We consider a static magnetic field $B_z=0.50 T$.  }
    \label{fig:LMC_vs_tx.png}
\end{figure*}


Having discussed the response with regards to  the magnetic field, we now analyze the angular dependence of LMOC and TMOC. In the absence of OMM and for linearly polarized light along the $z$-axis, the angular dependence profiles of the LMOC and TMOC are found to be $\delta\sigma_{zz}(\gamma)\propto B^2 \cos^2(\pi/2 -\gamma)= B^2  \sin^2\gamma$ and $\sigma_{xz}(\gamma)\propto B^2  \sin (\pi-2\gamma)=B^2  \sin 2\gamma$, consistent with the earlier works on magnetoconductivities~\cite{nandy2017chiral,ahmad2021longitudinal,nag2021magneto}. The  magnetotransport coefficients, associated with chiral anomaly and Berry curvature physics, arise from the quadratic-in-$B$ sector and are responsible for the $\sin^{2}\gamma$ and $\sin2\gamma$ angular profiles. Interestingly, increasing driving frequency reduces the magnitudes without altering the angular dependence of the conductivity for both LMOC and TMOC  as shown in Figs.~\ref{fig:LMC_PHC_vs_gm.png} (a) and (c), respectively, in the absence of OMM.  Further, the inclusion of the OMM changes the behavior of the conductivity qualitatively as shown in Fig.~\ref{fig:LMC_PHC_vs_gm.png}(b)
where LMOC acquires negative values in certain windows of $\gamma$, and the magnitude also does not decrease with $\omega$. On the other hand,  
for TMOC, the suppression of magnitude is more prominent in the presence of OMM as shown in Fig.~\ref{fig:LMC_PHC_vs_gm.png} (d).   In the case of circularly polarized light,
the angular dependence of the conductivities retains the same characteristic forms $\mathrm{LMOC}\propto\sin^{2}\gamma$ and $\mathrm{TMOC}\propto\sin2\gamma$. Note that this is in accordance with experimental observations in Refs.~\cite{liang2015ultrahigh,xiong2015evidence}.

We now discuss the origin behind the shift in LMOC, observed in  Fig.~\ref{fig:LMC_PHC_vs_gm.png}(b) 
in the presence of OMM. The OMM modifies the band dispersion, $\epsilon_{\mathbf{k}}^{\chi}=\epsilon_{\mathbf{k}}-\mathbf{m}_{\mathbf{k}}^{\chi}\!\cdot\!\mathbf{B}$, which generates corrections to the band velocity and density of states that are linear {as well as non-linear} in the magnetic field~\cite{gao2022suppression,gupta2024magneto,onsager1931reciprocal,sundaram1999wave}. As a result, the LMOC receives a {leading} linear-$B$ correction from OMM in addition to anomaly-driven $B^2$  contribution; such a term survives even after subtracting $\sigma_{zz}(B=0)$ from $\sigma_{zz}(B\ne 0)$~\cite{gao2022suppression,gupta2024magneto,sundaram1999wave}. This linear correction term in $B$ due to OMM is also expected to be present for TMOC as well.  {An explicit calculation shows that OMM can cause $B^2E$ contributions in current. This is essentially due to the fact that OMM
introduces energy and velocity corrections $\propto B$. The OMM results in a further correction to the non-equilibrium Fermi distribution function  $g^{\chi}_{\bf k} = -e \left( \frac{\partial f_{0}^{\chi}}{\partial {\color{black}\epsilon_{\bf{k}}^\chi}}\right) E_z~\Lambda^{\chi}_{{z,\bf k} }$  in terms of linear and non-linear powers of $B$. The same holds true for the equation of motion of $\dot{\mathbf{r}}_{\mathbf{k}}(t)$  once OMM contributions are taken into account. As a result, OMM  would yield an explicit $B^2 E$ contribution  while there will be implicit contributions of magnetic field coming from $\tau $ and coefficients $a,b,c,d$, see Appendix \ref{app-new}. }  
For experimentally relevant magnetic fields of order $B\sim1~\mathrm{T}$, the OMM contribution is not negligible and leads to a measurable offset in the angular dependence of both LMOC and TMOC. Importantly, the shift is predominantly visible in the case of  circularly polarized light for both LMOC and TMOC, see Figs.~\ref{fig:phc_vs_gm.png} and ~\ref{fig:LMC_vs_gm.png} in Appendix \ref{appd}.

\subsection{Magneto-optical response for tilted Weyl cones}
\label{res2}
We extend our analysis of the magneto-optical response to the tilted Weyl cones, considering the  Hamiltonian Eq. (\ref{Hamiltonian})  with $0<t^{\chi}_{x,z}\le 1 $. 
The  tilt generates the anisotropy in the Fermi surface, modifies the group velocity, thereby controlling the magneto-transport by varying tilt through strain, pressure or symmetry-breaking fields~\cite{das2019linear,ahmad2023longitudinal,cortijo2016linear}. We show below the variations of LMOC and TMOC with the above tilt terms which can be parallel or perpendicular to the static magnetic field with $\gamma=\pi/2$ such that ${\mathbf B}=(0,0,B_z)$.

\begin{figure*}
    \centering
    \includegraphics[width=1.95\columnwidth]{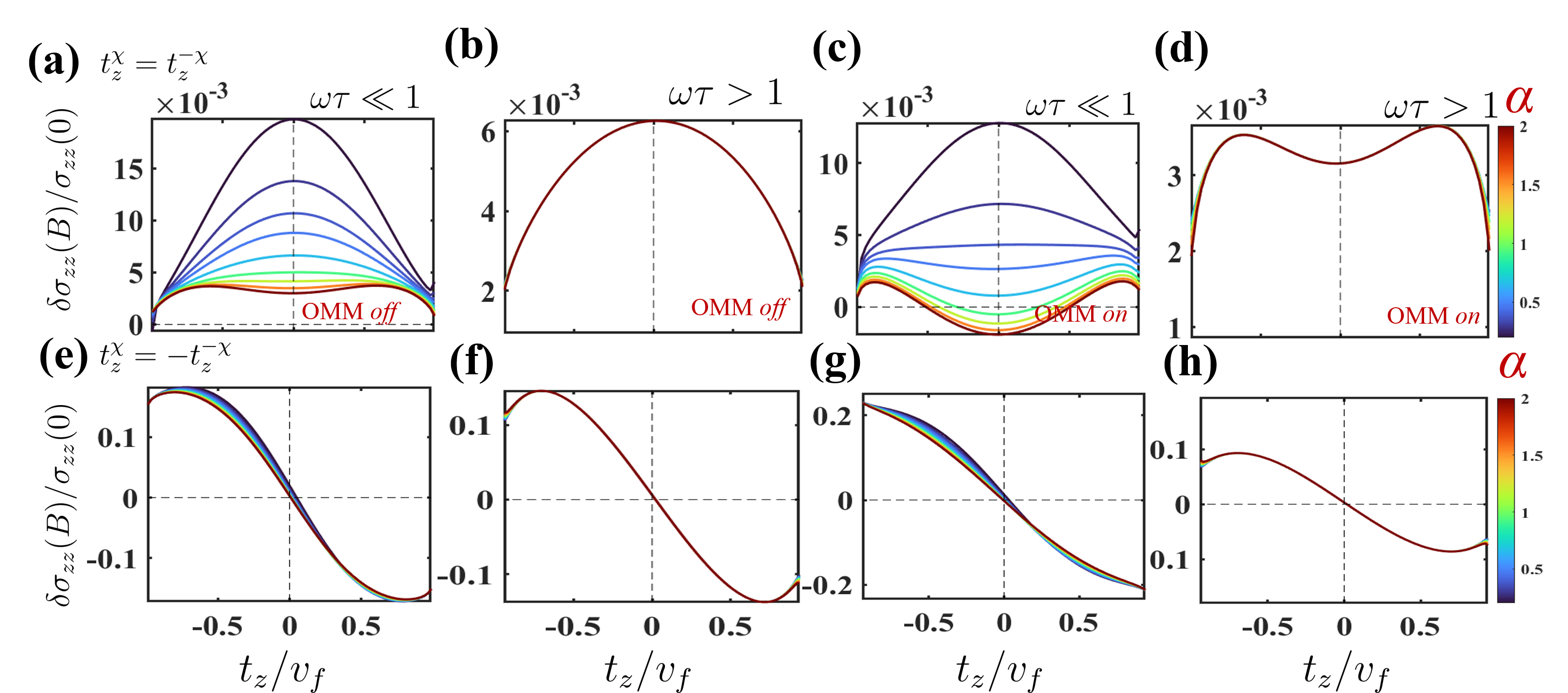}
    \caption{ We repeat Fig. \ref{fig:LMC_vs_tx.png}  for the tilt direction parallel to the applied magnetic field i.e., $t_z^{\chi} || B_z$ under linearly polarized electromagnetic radiation with $E_z$. We examine two situations with identical and opposite tilt orientations of the Weyl cones with $t_z^{\chi}= t_z^{-\chi}$ and $t_z^{\chi}= -t_z^{-\chi}$ in (a,b,c,d) and (e,f,g,h), respectively while the sign reversal of LMOC is observed for latter case without OMM and even for weak intervalley scattering.   
    We consider a static magnetic field $B_z=0.50 T$.}
    \label{fig:LMC_vs_tz.png}
\end{figure*}


We first examine the situation with the transverse tilt $t_x^\chi$  which is perpendicular to the magnetic field $\textbf{B}=(0,0,B_z)$ i.e., $t_x^\chi \perp B_z$. 
For this configuration, the LMOC is plotted in Figs.~\ref{fig:LMC_vs_tx.png} (a,b,c,d), and (e,f,g,h) for identically tilted Weyl cones with $t_x^{\chi}= t_x^{-\chi}$  and oppositely  with  $t_x^{\chi}= -t_x^{-\chi}$, respectively, keeping  \(t_z^{\chi}=0\). In the absence of OMM with weak intervalley scattering $\alpha \to 0$, the magnitude of the LMOC is suppressed with increasing tilt magnitude, irrespective of its orientation, see Figs.~\ref{fig:LMC_vs_tx.png} (a,e). With increasing $\alpha$, the monotonic fall changes into a non-monotonic one where LMOC as a function of tilt magnitude increases more rapidly for oppositely tilted Weyl cones as compared to the identically tilted case. Importantly, the response remains symmetric about \(t_x^{\chi}=0\) for $t_x^{\chi}=\pm t_x^{-\chi}$ indicating the fact that the relative orientation of the tilt matters only. Interestingly,    for both the tilt orientations, there exists a finite optimal tilt strength where the response is maximum for strong intervalley scattering $\alpha \to 2$. 
{The non-monotonic tilt dependence of the LMOC can be understood from the decomposition $\sigma_{zz}=\sigma_{zz}^{d}+\sigma_{zz}^{h}+\sigma_{zz}^{a}+\sigma_{zz}^{b}+\sigma_{zz}^{c}$ presented in Appendix~\ref{App:Non-monotonic dependence of MOR on tilts $t_{x/z}$}. Tilt modifies the Fermi-surface geometry, carrier velocities, and angular weightage of the transport contributions. For parallel $z$-tilts in the low-frequency regime, $\delta\sigma_{zz}^{d}/\sigma(0)$ dominates at weak intervalley scattering and for values of  tilt, it yields a monotonic response. Increasing $\alpha$ suppresses this contribution more rapidly than the second leading contribution $\delta\sigma_{zz}^{h}/\sigma(0)$, which exhibits non-monotonic behavior at larger values of tilt. The competition between these contributions determines the overall profile and the emergence of an optimal tilt strength. Thus, momentum-dependent scattering and charge conservation jointly control their relative weights, and the scattering-dependent evolution of the non-monotonic LMOC is a direct consequence of their combined effect. 
In the high-frequency regime, the response becomes less sensitive to intervalley scattering, and the non-monotonic behavior is governed predominantly by $\delta\sigma_{zz}^{h}/\sigma(0)$ contribution. This feature is absent in the previous works where constant relaxation was inherently assumed ~\cite{gao2022suppression,gupta2024magneto,liu2021role,kundu2020magnetotransport,detassis2017collective,carbotte2016dirac}}. 

%
Therefore, there exist qualitative differences between the responses even in the weak ac limit for different tilt orientations. On the other hand, in the strong ac limit, the effect of intervalley scattering is suppressed resulting in an identical profile of of LMOC with the tilt strength, see Figs.~\ref{fig:LMC_vs_tx.png} (b,f).  However, similar to the weak limit,  there exists a finite optimal tilt strength where the response is maximum. The above analysis of the weak ac limit clearly suggests that
spherically symmetric Fermi surfaces, associated with untilted Weyl cones, contribute the most for weak intervalley scattering while the deformed Fermi surfaces, associated with titled Weyl cones,  contribute maximally for strong intervalley scattering.

We now repeat the same analysis in the presence of OMM as shown in   Figs.~\ref{fig:LMC_vs_tx.png} (c,d) and (g,h) for 
identically and oppositely tilted Weyl cones, respectively. While in the weak ac limit, the overall response remains qualitatively similar to that in the absence of OMM, the LMOC becomes negative in the strong intervalley scattering regime \cite{ahmad2021longitudinal,knoll2020negative}. In the strong ac limit, identically and oppositely tilted Weyl cones produce same response which is insensitive to $\alpha$ as expected. The response remains positive under the strong intervalley scattering irrespective of the OMM as seen in Figs.  ~\ref{fig:LMC_vs_tx.png} (b,d,f,h).

Having discussed the perpendicular tilt configuration, we now focus on the parallel one $t_z^\chi ||B_z$ in Fig.~\ref{fig:LMC_vs_tz.png} (a,b,c,d) and (e,f,g,h) for $t_z^{\chi}= t_z^{-\chi}$ and   $t_z^{\chi}= -t_z^{-\chi}$, respectively while keeping the perpendicular component zero $t_x^{\chi}=0$.
At the outset, one can comment that 
the parallel tilt configuration produces qualitatively distinct responses in the oppositely tilted case as compared to the  perpendicular tilt configuration, see Figs.  ~\ref{fig:LMC_vs_tx.png} (e,f,g,h) and \ref{fig:LMC_vs_tz.png} (e,f,g,h). On the other hand, there exist qualitative similarities when it comes to the identically tilted Weyl cones, see  Figs.  ~\ref{fig:LMC_vs_tx.png} (a,b,c,d) and \ref{fig:LMC_vs_tz.png} (a,b,c,d). In the strong ac regime, the LMOC becomes monotonic and  insensitive to $\alpha$, displaying a quadratic scaling with $t_z^\chi$, see Fig.~\ref{fig:LMC_vs_tz.png}(b). This is distinct as compared to its previous counterpart which shows monotonic behavior for $t_x^\chi \perp B_z$, see Fig.~\ref{fig:LMC_vs_tx.png}(b). 
Therefore, the effect of the tilt orientation, whether it is identical or opposite, plays a significant role in determining the response when the direction of the tilt is parallel to the magnetic field. Interestingly, the LMOC becomes negative in the absence of OMM and changes its sign with the sign of the tilt for $t_z^{\chi}= -t_z^{-\chi}$, see Figs. \ref{fig:LMC_vs_tz.png} (e,f). Surprisingly, the quadratic behavior of 
LMOC with $t_z$ in the identically tilted case is replaced with linear behavior in the oppositely tilted case, see Figs. \ref{fig:LMC_vs_tz.png} (b,f).  The response remains almost unaltered in the presence of  OMM signifying the fact that the Fermi surface is maximally modified when the tilt is parallel to magnetic field   irrespective of OMM. For $t_z^{\chi}= -t_z^{-\chi}$, the LMOC for the untilted Weyl cone is vanishingly small and the response builds up as the tilt increases. This  is universally  
observed in weak and strong ac limit with  weak as well as strong intervalley coupling. This is   
markedly different from all the other cases, emphasizing the importance of tilt direction and its orientation on the response.

Given the above response profile with and without OMM for parallel tilt, one can comment that the  inclusion of OMM  qualitatively alters the behavior of LMOC. In the weak ac limit, OMM corrections introduce additional Berry curvature-mediated contributions to the semiclassical equations of motion, resulting in a sign reversal of the LMOC upon increasing $\alpha$, see Fig.~\ref{fig:LMC_vs_tz.png}(c). This highlights the crucial role of the self-rotation of Bloch wave packets in modifying magnetotransport under strong intervalley scattering. In the strong ac limit, unlike the OMM-free case where the response remains monotonic, the LMOC develops a pronounced non-monotonic dependence on $t_z$ while being insensitive to intervalley scattering strength,
as shown in Fig.~\ref{fig:LMC_vs_tz.png}(d). This underscores the frequency-dependent competition between dynamical contributions, arising from the ac electric field, and geometric contributions, caused by  Berry curvature and OMM. The inference can also be drawn for perpendicular tilt direction, as shown in Figs. \ref{fig:LMC_vs_tx.png}(c,d,g,h)
where the tilt orientation does not affect the response significantly. By contrast,   
for opposite tilt orientations with parallel magnetic field, $t_z^{\chi}= - t_z^{-\chi} || B_z$, the LMOC exhibits non-monotonic and asymmetric behavior with respect to the sign of the tilt magnitude which exhibits 
markedly distinct features, as shown in Figs.~\ref{fig:LMC_vs_tz.png}(e,f,g,h).
The Fermi surface gets maximally anisotropic for the parallel tilt direction
where the opposite tilt orientation minimizes the possibility of intervalley scattering even in the absence of OMM.

The dependence on $\alpha$ is comparatively weak and becomes negligible in the strong ac limit without OMM, where the response is dominated by intrinsic dynamical effects rather than intervalley relaxation, see Figs.~\ref{fig:LMC_vs_tz.png}(e,f). Incorporating OMM corrections once again leads to quantitative changes in the strong ac limit: at large intervalley scattering strengths, the LMOC displays an approximately linear dependence on $t_z$ within a narrow tilt window ranging from negative to positive value of tilt, see Figs.~\ref{fig:LMC_vs_tz.png}(h). To be precise,  the linear behavior in the strong ac limit is observed irrespective of OMM.  This observation clarifies that previously reported linear scaling behaviors are restricted to a limited parameter regime~\cite{gao2022suppression,gupta2024magneto,liu2021role,kundu2020magnetotransport,detassis2017collective,carbotte2016dirac}. Under circularly polarized illumination,  the overall trends remain qualitatively similar to those obtained for linearly polarized light. Although helicity induces chirality-dependent changes via Berry curvature coupling to the electromagnetic field, the main features—such as the response in weak and strong AC regimes, across different intervalley scattering strengths, and with OMM—remain qualitatively unchanged. Hence, for brevity, the corresponding circular polarization results are not shown explicitly.

\section{Conclusion}
\label{sec:conclusion}


We investigate the linear magneto-optical transport in WSMs within the semiclassical framework, focusing on the interplay of electromagnetic radiation, Berry curvature,  intervalley scattering,  momentum-dependent scattering time, charge conservation, OMM, and tilting of Weyl cones.  For untilted Weyl cones,
the weak ac limit $\omega\tau\ll1$ exhibits the well-known sign reversal of the LMOC for strong intervalley scattering. The degree of sign-reversal is reduced and eventually disappears in strong ac limit $\omega\tau\gtrsim1$, where the LMOC remains strictly positive. This clearly indicates that increasing the driving frequency suppresses the buildup of chiral imbalance over a cycle of electromagnetic radiation, thereby weakening the effect of intervalley scattering.  The sign reversal of LMOC is attributed to the presence of OMM which further modifies the angular dependence of magnetotransport. Interestingly, LMOC  receive a linear-in-$B$ background contribution from OMM  while preserving the characteristic angular profiles $\sin^{2}\gamma$ and $\sin2\gamma$ behavior, respectively, associated with anomaly-induced $B^2$ term. We extend our analysis to circularly polarized light where we find qualitatively similar results to that of linearly polarized light.

We explore the above responses for the tilted  Weyl cones which allow us to investigate the effects of magnitude, direction, and orientation of the tilt on LMOC and TMOC. Considering  tilt  \(t_x^{\chi}=\pm t_x^{-\chi}\) in the perpendicular direction with respect to magnetic field $B_z$ [$+(-)$ refers to identical (opposite) tilt orientation], we find that the LMC exhibits a  non-monotonic dependence on the tilt strength while the response is symmetric with positive and negative values of tilt. In the absence of OMM,  the LMOC  does not show sign reversal, while OMM restores it only in the weak ac regime. On the other hand, the intervalley 
scattering causes the LMOC signal to diminish in the weak ac limit irrespective of OMM. Interestingly, the opposite orientation of tilt can quantitatively  modify the non-monotonic profile with tilt strength while keeping the symmetry around the zero tilt unaltered. For a tilt parallel to the magnetic field direction, $t_z^{\chi}=\pm t_z^{-\chi}$ with $t_x^{\chi}=0$, the LMOC acquires a profile that substantially depends on frequency and intervalley scattering. In the weak ac limit, LMOC exhibits a non-monotonic dependence on the tilt magnitude with large intervalley scattering. In contrast, in strong ac regime, the LMOC becomes monotonic, essentially independent of $\alpha$, and scales quadratically with the tilt magnitude. Similar to the earlier tilt direction, OMM causes the LMOC to reverse its sign from positive to negative for large intervalley scattering.  
Interestingly, the opposite orientation of tilt qualitatively alters the non-monotonic profile which is transformed into a nearly monotonic without any symmetry around the zero tilt.  For oppositely tilted Weyl cones,  the LMOC depends only weakly on the intervalley scattering and becomes nearly independent of the above  in the strong ac regime. The same effect continues to survive in the presence of OMM as well.    It is important to note that all the above observations  qualitatively  persist under circularly polarization, indicating the fact  that the essential features originate from the interplay of longitudinal tilt, intervalley scattering, and OMM corrections.

{In particular, our work reveals several features that are absent in the dc limit and have not been discussed in earlier studies. These include the following: the frequency-driven suppression of intervalley-scattering effects, leading to the disappearance of the sign reversal of the LMOC in the strong ac regime;  the OMM is found to modify both the magnitude and angular dependence of magneto-optical responses in weak as well as strong ac limits;  the LMOC 
changes non-monotonically with the tilt leading to distinct frequency-dependent transport signatures associated with different tilt directions and orientations of the Weyl cones.} Our results thus demonstrate that frequency-dependent LMOC measurements provide a sensitive probe of chiral anomaly physics, intervalley dynamics, OMM effects, and both the magnitude and orientation of tilt in the  Weyl cones, offering clear experimental signatures accessible through magneto-optical transport ranging from the MHz to the THz regime. In the future, one can study the effect of OMM in the non-linear magneto-optical response.

\section{Acknowledgements}
A.A. and T.N. acknowledge the NFSG- NFSG/HYD/2023/H0911 from BITS Pilani. S.~N. acknowledges financial support from Anusandhan National Research Foundation (ANRF), Government of India via the Prime Minister's Early Career Research Grant: ANRF/ECRG/2024/005947/PMS. P.~B. acknowledges financial support from the Anusandhan National Research Foundation under project number SUR/2022/000289. T.N. acknowledges the Advanced Research Grant (ARG) from Anusandhan National Research Foundation Grant No. ANRF/ARG/2025/003163/PS. A.A. is grateful to Binita Malik for fruitful discussions and IIT Mandi for providing computing facilities. 

\appendix

\section{Generalization to circularly polarized light  }
\label{sec:Generalization to circularly polarised light}

We now extend the above semiclassical Boltzmann formalism to the case of circularly polarised light~\cite{sodemann2015quantum}. All equations of motion, collision integrals, and Boltzmann equations derived in the Sec.~\ref{sec:MBT} remain unchanged. The only modification arises from the structure of the time-dependent electric field. For circularly polarised light confined to the $x$--$z$ plane, the electric field is written in a compact complex representation as
\begin{equation}
\mathbf{E}(t)
=
E\,\frac{1}{\sqrt{2}}
\left(i\eta\,
\hat{x} + \hat{z}
\right)
e^{-i\omega t},
\label{Eq:CPL_field}
\end{equation}
where $\eta=\pm1$ denotes the helicity of the light. The physical electric field 
is obtained by taking the real part of Eq.~\eqref{Eq:CPL_field}. Using the general ansatz introduced in Eq.~\eqref{Eq:g1}, 
The first-order correction to the distribution function becomes
\begin{align}
g^{\chi}_{\mathbf{k}}
&= -e \left(
\mathbf{E}\cdot\mathbf{\Lambda}^{\chi}_{\mathbf{k}}
\frac{\partial}{\partial\epsilon_{\mathbf{k}}}
\right) f_0 \nonumber\\
&= -\frac{e}{\sqrt{2}}
\left(
i\eta\, E_x\Lambda^{\chi}_{x,{\mathbf{k}}}
+ E_z \Lambda^{\chi}_{z,{\mathbf{k}}}
\right)
\left(\frac{\partial f_0}{\partial\epsilon_{\mathbf{k}}}\right).
\label{Eq:g1_CPL}
\end{align}
Substituting Eq.~\eqref{Eq:CPL_field} into the definition of 
$\mathcal{A}^{\chi}_{\mathbf{k}}$ appearing in 
Eq.~\eqref{Eq_boltz21}, we obtain
\begin{align}
\mathcal{A}^{\chi}_{\mathbf{k}}
&=
\left(
\mathbf{E}
+ \frac{e}{\hbar}(\mathbf{E}\cdot\mathbf{B})\boldsymbol{\Omega}^{\chi}
\right)\cdot\mathbf{v}^{\chi}_{\mathbf{k}}
\nonumber\\
&=
\frac{1}{\sqrt{2}}
\left( i\eta\, E_x
v^{\chi}_{x,{\mathbf{k}}}
+E_z v^{\chi}_{z,{\mathbf{k}}}
\right)
\nonumber\\
&+\frac{eB}{\sqrt{2}\hbar}
(\boldsymbol{\Omega}^{\chi}_{\mathbf{k}}\cdot \mathbf{v}^{\chi}_{\mathbf{k}})
\left(i\eta\cos\gamma E_x
+ \sin\gamma E_z
\right).
\label{Eq:Achi_CPL}
\end{align}
The expression for the current density remains identical to 
Eq.~\eqref{Boltzman_final} for $\Lambda^{\chi}_{x,\textbf{k}}~
\&~~\Lambda^{\chi}_{z,\textbf{k}}$. The linear current response in the first-order of $E$,  can be written as
\begin{align}
J_{\alpha}
=
\sum_{\beta}
\sigma_{\alpha\beta}(\omega)\,
E_{\beta},
\qquad
E_{\beta}
= (E_x,0,E_z).
\end{align}
Note that the circularly polarized light   can induce currents along $\hat{z}$ and $\hat{x}$-directions while the electric field is also present along $\hat{x}$-direction, resulting in the emergence of $\sigma_{xx}$. To be precise, $J_x(t)= \sigma_{xx}(\omega) E_x (t)+ \sigma_{xz}(\omega) E_z (t)$ and $J_z(t)= \sigma_{zx} (\omega) E_x (t)+ \sigma_{zz}(\omega) E_z (t)$ where $\sigma_{xx} (\omega)$  and $\sigma_{zx}(\omega)$ both  involve $\Lambda^{\chi}_{x, {\bf k}}(\omega)$. Note that $\sigma_{xx} (\omega)$ and $\sigma_{zx}(\omega)$  conductivities can only be present for circularly polarized light.

\section{Zero magnetic field conductivity under linearly polarized light}
\label{appB}
For $B=0$, electric field along $z$-direction, under relaxation time approximation i.e., momentum-independent relaxation time,  $\widetilde{\tau}^{\chi}_{\mathbf{k}}(\omega)=\widetilde{\tau}^{\chi}_{c}(\omega)$,  the ansatz reduces to $\Lambda^{\chi}_{z,\mathbf{k}} = v^{\chi}_{z}~\widetilde{\tau}^{\chi}_{c}(\omega)$ where all the correction terms $d^{\pm}, a^{\pm }, b^{\pm }, c^{\pm }$ vanishes. This leads to the following form of conductivities,
\begin{align}
    \sigma_{xz} &=\sum_{\chi}\frac{-e^2~\widetilde{\tau}^{\chi}_{c}(\omega)}{\hbar}  \int\frac{d^3\mathbf{k}}{(2\pi)^3} \left[\hbar v_{x,\textbf{k}}^{\chi} v^{\chi}_{z,\textbf{k}} \left( \frac{\partial f^\chi_{0}}{\partial {\epsilon_{\textbf{k}}}}\right) + \Omega_{y,\textbf{k}}^\chi f^\chi_{0} \right],\nonumber\\
    \sigma_{yz} &=\sum_{\chi}\frac{-e^2\widetilde{\tau}^{\chi}_{c}(\omega)}{\hbar}\int\frac{d^3\mathbf{k}}{(2\pi)^3}\left[\hbar v_{y,\textbf{k}}^{\chi} v^{\chi}_{z,\textbf{k}} \left( \frac{\partial f^\chi_{0}}{\partial {\epsilon_{\textbf{k}}}}\right) - \Omega_{x,\textbf{k}}^\chi f^\chi_{0} \right],\nonumber\\ 
    \sigma_{zz} &=\sum_{\chi}\frac{-e^2~\widetilde{\tau}^{\chi}_{c}(\omega)}{\hbar} \int\frac{d^3\mathbf{k}}{(2\pi)^3}\left[\hbar v_{z,\textbf{k}}^{\chi} v^{\chi}_{z,\textbf{k}}\left( \frac{\partial f^\chi_{0}}{\partial {\epsilon_{\textbf{k}}}}\right)\right].
\end{align}
In the above equation, we have used, $g^{\chi}_{\textbf{k}} = -e \left( \frac{\partial f_{0}}{\partial {\epsilon_{\textbf{k}}}}\right) E ~v^{\chi}_{z,\textbf{k}}~\widetilde{\tau}^{\chi}_{c}(\omega)$, with, $1/\widetilde{\tau}^{\chi}_{c}(\omega) = (\frac{1}{\tau^{\chi}_{c}} -i\omega)$. The above equations are in accordance with the one derived in Refs.~\cite{sodemann2015quantum,zhang2023higher,gao2022suppression,gupta2024magneto,dagnino2025non,prabhat2026magneto}. The first-order current is  given by $J_{\mu}(t)=\mathrm{Re}\!\left[\sigma_{\mu\alpha}(\omega)\,
E_{\alpha}\,e^{i\omega t}\right]$,
with
\begin{equation}
\sigma_{\mu\alpha}(\omega)
=\frac{e^{2}}{\hbar}\sum_{\chi}\int_{\mathbf{k}} f_{0}
\left(
\frac{\partial_{\mu}\partial_{\alpha}\epsilon_{\mathbf{k}}}
{\hbar\widetilde{\omega}}
-\epsilon_{\mu\alpha\beta}\Omega_{\beta,\textbf{k}}
\right),
\end{equation}
and, $ \widetilde{\omega}=-i\omega+1/\tau$. Here, $\int_{\mathbf{k}} =\int\frac{d^3\mathbf{k}}{(2\pi)^3}$. Using the band velocity $v_{\mu,\textbf{k}}=\frac{1}{\hbar}\partial_{\mu}\epsilon_{\mathbf{k}}$, the second derivative of the dispersion can be written as $\partial_{\mu}\partial_{\alpha}\epsilon_{\mathbf{k}}
=\hbar^{2}\,\partial_{k_{\alpha}} v_{\mu,\textbf{k}}$. Integrating by parts in momentum space and using $\partial_{k_{\alpha}} f^\chi_{0}
=\frac{\partial f^\chi_{0}}{\partial\epsilon_{\textbf{k}}}\,
\partial_{k_{\alpha}}\epsilon_{\mathbf{k}}
=\hbar v_{\alpha,\textbf{k}}\frac{\partial f^\chi_{0}}{\partial\epsilon_{\textbf{k}}}$, we obtain
\begin{equation}
\int_{\mathbf{k}} f_{0}\,\partial_{\mu}\partial_{\alpha}\epsilon_{\mathbf{k}}
=-\hbar^{2}\int_{\mathbf{k}} v_{\mu,\textbf{k}}v_{\alpha,\textbf{k}}
\left(\frac{\partial f^\chi_{0}}{\partial\epsilon_{\textbf{k}}}\right).
\end{equation}
Substituting this back into $\sigma_{\mu\alpha}$ gives
\begin{equation}
\sigma_{\mu\alpha}(\omega)
=\frac{e^{2}}{\hbar}\sum_{\chi}\int_{\mathbf{k}}
\left[
\hbar\, v_{\mu,\textbf{k}}v_{\alpha,\textbf{k}}\,
\widetilde{\tau}(\omega)
\left(\frac{\partial f^\chi_{0}}{\partial\epsilon_{\textbf{k}}}\right)
-\epsilon_{\mu\alpha\beta}\Omega_{\beta,\textbf{k}}\,f^\chi_{0}
\right],
\end{equation}
where $\widetilde{\tau}(\omega)=1/(-i\omega+1/\tau)$. Choosing $\mu=x$ and $\alpha=z$, and multiplying by $E_{z}$, we recover the linear current
\begin{equation}
J_{x}
=\frac{e^{2}}{\hbar}\sum_{\chi,\mathbf{k}}
\left[
\hbar v^{\chi}_{x,\textbf{k}}v^{\chi}_{z,\textbf{k}}\,
\widetilde{\tau}^{\chi}_{c}(\omega)
\left(\frac{\partial f^\chi_{0}}{\partial\epsilon_{\textbf{k}}}\right)E_z
+\Omega^{\chi}_{y,\textbf{k}}f^\chi_{0}E_z
\right].
\end{equation}
Similarly, one can verify other components of current as well. This provides a solid foundation of our conductivity expressions for a finite magnetic field as it correctly reproduces the zero magnetic field results. 


\begin{figure}
    \centering
    \includegraphics[width=.95\columnwidth]{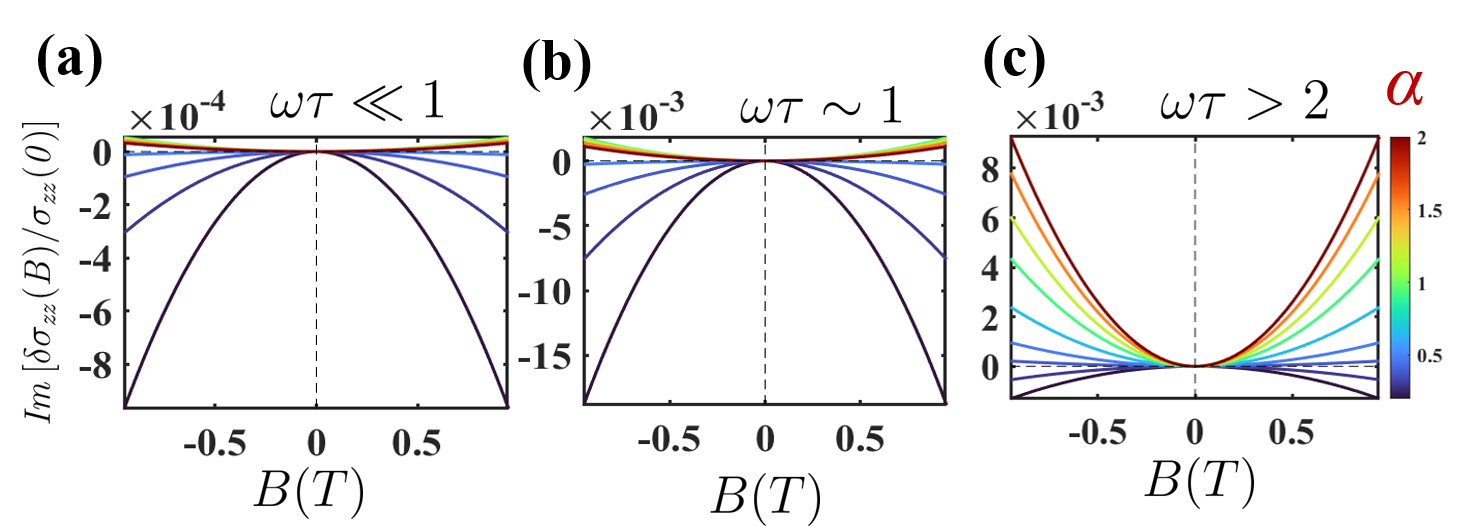}
    \caption{We show the imaginary part of LMOC $\delta \sigma_{zz}(B)/\sigma_{zz}(0)={\rm Im} [\sigma_{zz}(B\ne 0)/\sigma_{zz}(0)-1]$
    in (a,b,c) for weak, moderate and strong ac regimes, respectively under linearly polarized irradiation in the presence of OMM. The negative response even for strong intervalley scattering in weak ac limit, is a signature of the reactive part of LMOC.}
    \label{fig:Imag_LMC_vs_B_alp_vary1}
\end{figure}

{ 
\section{Effect of OMM on the conductivity}}
\label{app-new}

{ 
We make it explicitly clear by following calculation. Assuming the modified band velocity as ${\mathbf{v}} =\mathbf{v}_0 + B \mathbf{v}_B$, with $\mathbf{v}_0$ being bare band velocity and $B \mathbf{v}_B$ velocity due to OMM. The expansion of $D_{\mathbf k}$ will give:
\begin{align}
   \mathcal{D}_{\mathbf{k}}= 1 - \frac{e B \, \hat{\mathbf{z}} \cdot \boldsymbol{\Omega}}{\hbar}
+ \frac{e^2 B^2 \left( \hat{\mathbf{z}} \cdot \boldsymbol{\Omega} \right)^2}{\hbar^2}
\end{align}
For simplicity, we take $\mathbf{B} =B \hat{z}$ and $\mathbf{E} =E \hat{z}$. We will expand $\dot{\mathbf{r}}_{\mathbf{k}}$ as,
\begin{align}
\dot{\mathbf{r}}_{\mathbf{k}}(t) /\mathcal{D}_{\mathbf{k}} &=\mathbf{v}_0
+ B\,\mathbf{v}_B
+ \frac{e E}{\hbar}\,\hat{\mathbf{z}}\times\boldsymbol{\Omega}
+ \frac{e B}{\hbar}\,\hat{\mathbf{z}}
\left(\boldsymbol{\Omega}\cdot\mathbf{v}_0\right) \nonumber \\ 
&+ \frac{e B^2}{\hbar}\,\hat{\mathbf{z}}
\left(\boldsymbol{\Omega}\cdot\mathbf{v}_B\right).
\end{align}
We keep $D_{\mathbf k}$ as it is, as it will cancel in the current calculation. Under constant relaxation time approximation, the first-order deviation in the distribution function in the electric field is given by (from Eq. 10 of main text),
\begin{align}
\label{eqn:df}
    \delta f &= -eE \left(\frac{\partial f_0({\widetilde{\epsilon})}}{\partial \widetilde{\epsilon}}\right) \mathcal{D}_{\mathbf{k}} (v_z + \frac{eB}{\hbar} ({\mathbf{v}}\cdot {\mathbf{\Omega}})).
\end{align}    
To explicitly trace the different contributions in MOR, we will expand each term of the above equation in powers of E and B.\\
The field-induced correction to the band energy is assumed to be small, so that the modified energy can be written as $\widetilde{\epsilon} = \epsilon+\delta\epsilon$, where $\delta\epsilon = B\,\epsilon^{(1)}$, is the correction due to OMM. To be precise, 
$\delta\epsilon =-\mathbf{m}^{\chi}_{\mathbf{k}}\cdot\mathbf{B}$ and $\epsilon^{(1)} \equiv |\mathbf{m}^{\chi}_{\mathbf{k}}|$. Note that $\frac{1}{\hbar}\mathbf{\nabla} \widetilde{\epsilon}=\mathbf{v}=\mathbf{v}_0 + B \mathbf{v}_B$ with $\mathbf{\nabla} {\delta \epsilon}=B\mathbf{v}_B$.
Expanding the derivative of the Fermi-Dirac distribution about the unperturbed energy $\epsilon$ using a Taylor series, we obtain
\begin{align}
\frac{\partial f_0(\widetilde{\epsilon})}
{\partial\widetilde{\epsilon}}
=
\frac{\partial f_0}{\partial\epsilon}
+
\delta\epsilon
\frac{\partial^2f_0}{\partial\epsilon^2}
+
\frac{(\delta\epsilon)^2}{2}
\frac{\partial^3f_0}{\partial\epsilon^3}
+\mathcal{O}(\delta\epsilon^3).
\end{align}
Substituting $\delta\epsilon = B\,\epsilon^{(1)}$, gives $(\delta\epsilon)^2 = B^2(\epsilon^{(1)})^2$. Retaining terms up to second order in the magnetic field, we obtain
\begin{align}
\frac{\partial f_0(\widetilde{\epsilon})}
{\partial\widetilde{\epsilon}}
=
&
\frac{\partial f_0}{\partial\epsilon}
+
B\,\epsilon^{(1)}
\frac{\partial^2f_0}{\partial\epsilon^2}
+
B^2
\frac{(\epsilon^{(1)})^2}{2}
\frac{\partial^3f_0}{\partial\epsilon^3}
+\mathcal{O}(B^3).
\label{eq:df_expand}
\end{align} 
This brings Eq.~\eqref{eqn:df} as follows while keeping up to quadratic terms of magnetic field and $\hbar =1$,
\begin{align} 
\delta f &\simeq \frac{1}{2}\delta''(\epsilon-\epsilon_F)e(\epsilon^{(1)})^2v_{z0}B^2E\tau 
+\delta'(\epsilon-\epsilon_F)e \epsilon^{(1)} BE\tau  \nonumber \\
&\times 
\left(
v_{z0}
+v_{zB}B
+eB\boldsymbol{\Omega}\cdot\mathbf{v}_0
+ev_{zB}B^2\hat{\mathbf z}\cdot\boldsymbol{\Omega}
\right) \nonumber \\
&\times (1 - e B \, \hat{\mathbf{z}} \cdot \boldsymbol{\Omega}) 
\nonumber\\
&+\delta(\epsilon-\epsilon_F)eE\tau
\left(
v_{z0}
+v_{zB}B
+eB\boldsymbol{\Omega}\cdot\mathbf{v}_0
+ev_{zB}B^2\hat{\mathbf z}\cdot\boldsymbol{\Omega}
\right)\nonumber \\
& \times ( 1 - e B \, \hat{\mathbf{z}} \cdot \boldsymbol{\Omega}
+ e^2 B^2 \left( \hat{\mathbf{z}} \cdot \boldsymbol{\Omega} \right)^2).
\end{align}
Therefor total current upto $~EB^2$ is, $J^{EB^2} = \sum_{k} \delta f \dot{\mathbf{r}}_{\mathbf{k}}(t) /\mathcal{D}_{\bf{k}}$, explicitly,   
\begin{align}
    J^{EB^2} &\sim \frac{1}{2}\delta''(\epsilon-\epsilon_F)e(\epsilon^{(1)})^2
\mathbf{v}_{0}v_{z0}B^2E\tau
\nonumber\\
&
+\delta'(\epsilon-\epsilon_F)e\epsilon^{(1)}BE\tau
\big(
\mathbf{v}_{0}v_{z0}
+\mathbf{v}_{B}v_{z0}B
+\mathbf{v}_{0}v_{zB}B\nonumber \\
&+e\mathbf{v}_{0}B\boldsymbol{\Omega}\cdot\mathbf{v}_{0}
+e v_{z0}\hat{\mathbf z}B\boldsymbol{\Omega}\cdot\mathbf{v}_{0}
-e\mathbf{v}_{0}v_{z0}B\hat{\mathbf z}\cdot\boldsymbol{\Omega}
\big)
\nonumber\\
&
+\delta(\epsilon-\epsilon_F)eE\tau
\big(
\mathbf{v}_{0}v_{z0}
+\mathbf{v}_{B}v_{z0}B
+\mathbf{v}_{0}v_{zB}B
+\mathbf{v}_{B}v_{zB}B^2 \nonumber \\
&
+ev_{z0}\hat{\mathbf z}B\boldsymbol{\Omega}\cdot\mathbf{v}_{0}
+e\mathbf{v}_{B}B^2\boldsymbol{\Omega}\cdot\mathbf{v}_{0}
+ev_{zB}\hat{\mathbf z}B^2\boldsymbol{\Omega}\cdot\mathbf{v}_{0}
\nonumber\\
&
+e^2 \hat{\mathbf z}B^2
(\boldsymbol{\Omega}\cdot\mathbf{v}_{0})^2
+\mathbf{v}_{0}B^2\boldsymbol{\Omega}\cdot\mathbf{v}_{B}
+ev_{z0}\hat{\mathbf z}B^2\boldsymbol{\Omega}\cdot\mathbf{v}_{B}
\nonumber\\
&
-e\mathbf{v}_{0}v_{z0}B\hat{\mathbf z}\cdot\boldsymbol{\Omega}
-\mathbf{v}_{B}v_{z0}B^2\hat{\mathbf z}\cdot\boldsymbol{\Omega}
+e \mathbf{v}_{0}v_{zB}B^2\hat{\mathbf z}\cdot\boldsymbol{\Omega}
\nonumber\\
&
-e^2\mathbf{v}_{0}B^2
(\boldsymbol{\Omega}\cdot\mathbf{v}_{0})
(\hat{\mathbf z}\cdot\boldsymbol{\Omega})
-e^2v_{z0}\hat{\mathbf z}B^2
(\boldsymbol{\Omega}\cdot\mathbf{v}_{0})
(\hat{\mathbf z}\cdot\boldsymbol{\Omega}) \nonumber \\
&- eB^2 v_{z0} v_{zB}  \hat{\mathbf z}\cdot\boldsymbol{\Omega} +e^2\mathbf{v}_{0}v_{z0}B^2
(\hat{\mathbf z}\cdot\boldsymbol{\Omega})^2 +e\mathbf{v}_{0}B\boldsymbol{\Omega}\cdot\mathbf{v}_{0}
\big).
\end{align}
Now the terms which contribute to the current due to OMM (terms having either $\epsilon_1$ and/or $\mathbf{v}_B$ and/or $v_{zB}$),
\begin{align}
J^{EB^2}_{\rm OMM} &=\frac{1}{2}\delta''(\epsilon-\epsilon_F)e(\epsilon^{(1)})^2
\mathbf{v}_{0}v_{z0}B^2E\tau
\nonumber\\
&+\delta'(\epsilon-\epsilon_F)e\epsilon^{(1)}BE\tau
\big(
\mathbf{v}_{0}v_{z0}
+\mathbf{v}_{B}v_{z0}B
+\mathbf{v}_{0}v_{zB}B \nonumber \\
&+e\mathbf{v}_{0}B\boldsymbol{\Omega}\cdot\mathbf{v}_{0}
+ev_{z0}\hat{\mathbf z}B\boldsymbol{\Omega}\cdot\mathbf{v}_{0}
-e\mathbf{v}_{0}v_{z0}B\hat{\mathbf z}\cdot\boldsymbol{\Omega}
\big)
\nonumber\\
&
+\delta(\epsilon-\epsilon_F)eE\tau
\big(
\mathbf{v}_{B}v_{z0}B
+\mathbf{v}_{0}v_{zB}B
+\mathbf{v}_{B}v_{zB}B^2\nonumber \\
&+e\mathbf{v}_{B}B^2\boldsymbol{\Omega}\cdot\mathbf{v}_{0}
+ev_{zB}\hat{\mathbf z}B^2\boldsymbol{\Omega}\cdot\mathbf{v}_{0}
\nonumber\\
&
+\mathbf{v}_{0}B^2\boldsymbol{\Omega}\cdot\mathbf{v}_{B}
+ev_{z0}\hat{\mathbf z}B^2\boldsymbol{\Omega}\cdot\mathbf{v}_{B}
-\mathbf{v}_{B}v_{z0}B^2\hat{\mathbf z}\cdot\boldsymbol{\Omega} \nonumber \\
&+e\mathbf{v}_{0}v_{zB}B^2\hat{\mathbf z}\cdot\boldsymbol{\Omega} -eB^2 v_{z0} v_{zB}  \hat{\mathbf z}\cdot\boldsymbol{\Omega} \big).
\end{align}
From the above equation, it is clear that OMM can give linear as well as non-linear in $B$ contribution to the current.\\

One can connect the above discussion with the momentum-dependent scattering time formalism. We know that the deviation from the equilibrium Fermi distribution function is given by  $g^{\chi}_{\bf k} = -e \left( \frac{\partial f_{0}^{\chi}}{\partial {\color{black}\epsilon_{\bf{k}}^\chi}}\right) E_z~\Lambda^{\chi}_{{z,\bf k} }$ with the ansatz $\Lambda^{\chi}_{z,\mathbf{k}}=\big[d^{\chi}-h^{\chi}_{z,\mathbf{k}} + a^{\chi}\cos{\theta} +b^{\chi}\sin{\theta}\cos{\phi}+c^{\chi}\sin{\theta}\sin{\phi} \big] {\widetilde{\tau}^{\chi}_{\textbf{k}}(\theta,\phi)}$, where $a^\chi$, $b^\chi$, $c^{\chi}$ and  $d^\chi$ are the unknown coefficients having the dimension of velocity.  Here, $h^{\chi}_{z,\bf{k}}(\theta,\phi)=\mathcal{D}^{\chi}_{\mathbf{k}}[v^{\chi}_{z,\textbf{k}}+\frac{eB_z}{\hbar}(\mathbf{\Omega}^{\chi}_{k}\cdot \mathbf{v}^{\chi}_{\mathbf{k}})]$ with $B_z=B\sin{\gamma}$. Therefore, the expansion of $\mathcal{D}^{\chi}_{\mathbf{k}}$ and band velocity ${\mathbf{v}}$ in terms of magnetic field $B$ remains valid. This would yield explicit $B^2 E$ contribution from OMM while there will be implicit contributions coming from $\tau $ and coefficients $a,b,c,d$.\\  

}

\begin{figure}
    \centering
  \includegraphics[width=.95\columnwidth]{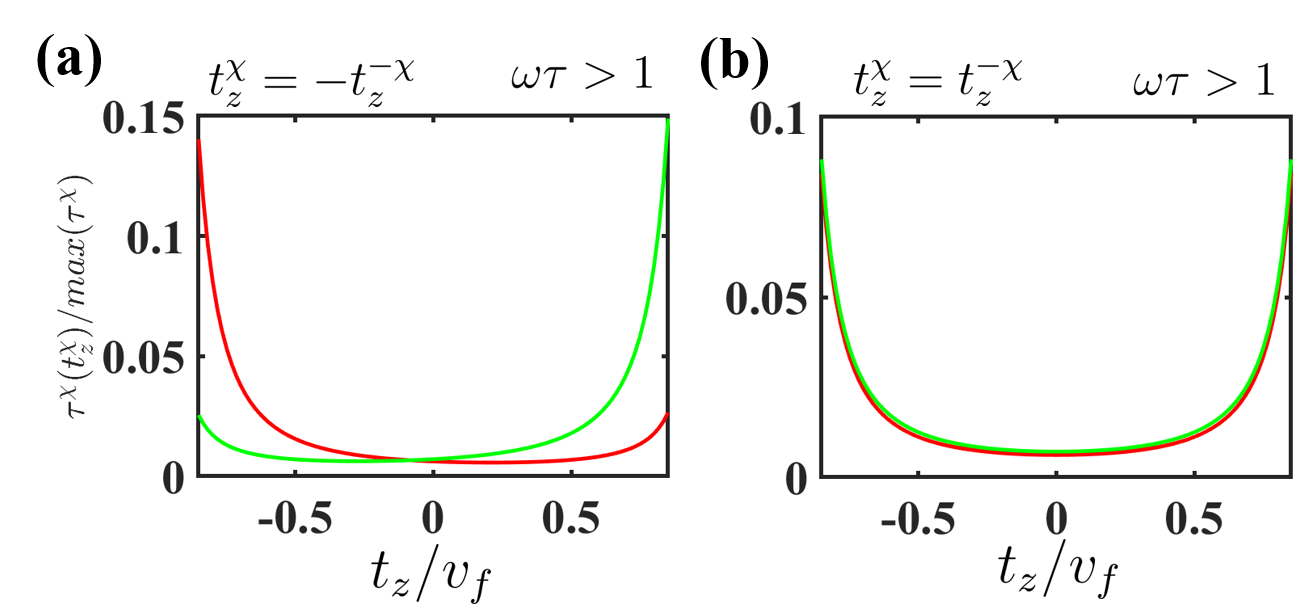}
    \caption{{ Chiral relaxation time $\tau^{\chi}(\theta,\phi)$ as a function of tilt $t_z/v_F$ at a fixed value of $\theta$ and $\phi$. Red and green colors denote the relaxation time, associated with positive and negative chirality indices, respectively.  (a,b) represent opposite and parallel tilt configurations, respectively.} }
\label{fig:tau_vs_tz_OMM1_opp_and_same}
\end{figure}
\begin{figure}
    \centering
  \includegraphics[width=.95\columnwidth]{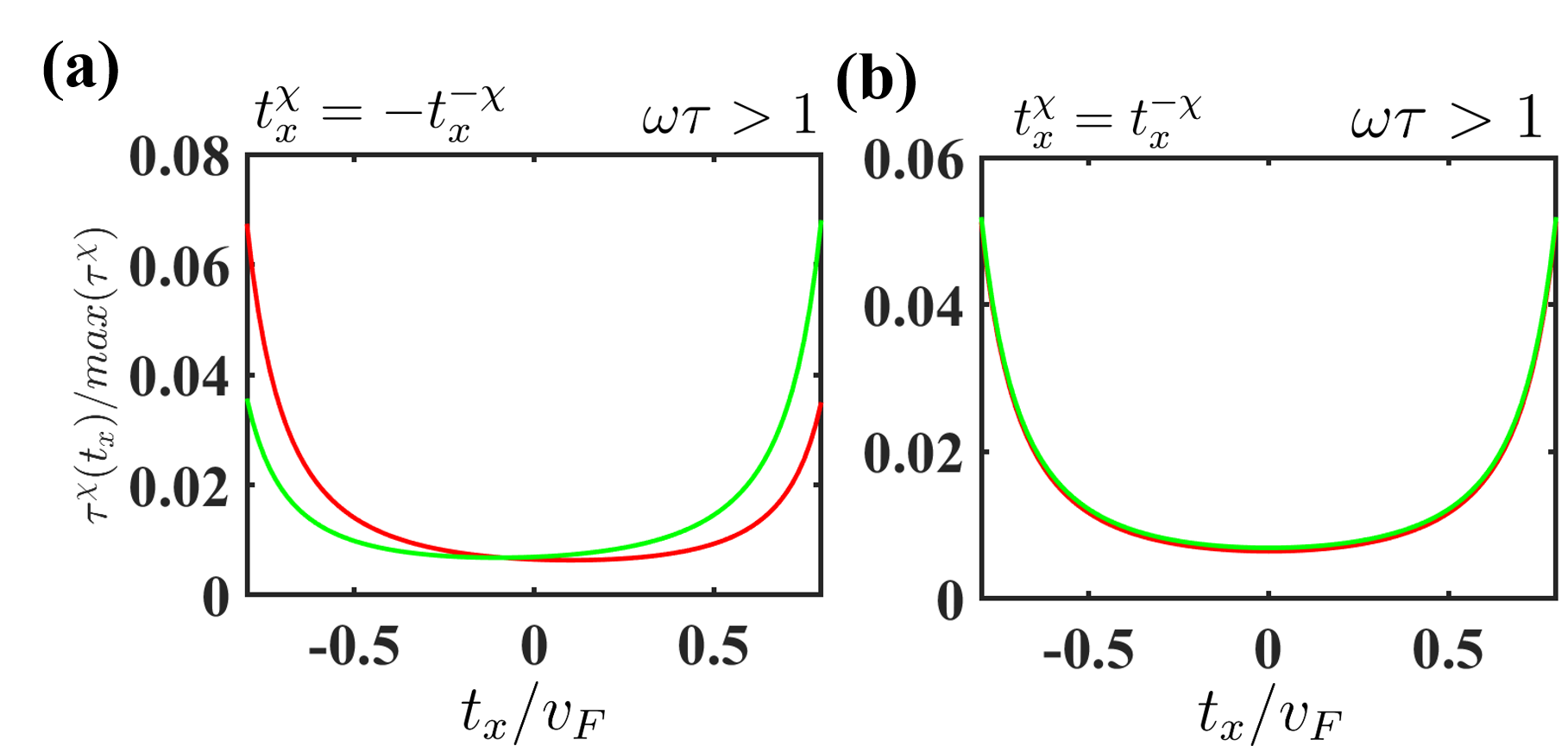}
    \caption{{ Chiral relaxation time $\tau^{\chi}(\theta,\phi)$ as a function of tilt $t_x/v_F$ at a fixed value of $\theta$ and $\phi$. Red and green colors denote the relaxation time, associated with positive and negative chirality indices, respectively.  (a,b) represent opposite and parallel tilt configurations, respectively.  }}
\label{fig:OMM1_tau_vs_tx_oppo_and_same}
\end{figure}

{ 
\section{Tilt-dependence of the relaxation time}
\label{app-new1}}

{ 
A finite tilt deforms the constant-energy surface and modifies both the group velocity ($\sim v_F t_{x,z} $) and the scattering phase space. Consequently, the chiral relaxation time $\tau^{\chi}(\theta,\phi)$ can be tilt dependent. For the confirmation, we plot the chiral relaxation time under parallel tilt configuration $t^{\chi}_z || B_z$ in the Figs.~\ref{fig:tau_vs_tz_OMM1_opp_and_same} (a,b)  for the opposite ($t^{\chi}_z = - t^{-\chi}_z $)  and same ($t^{\chi}_z =  t^{-\chi}_z $) tilt orientation, respectively, with positive (red) and negative (green) chiralities.
We repeat the same analysis for the perpendicular tilt configuration $t^{\chi}_x \perp B_z$, where we find identical results, see Figs.~\ref{fig:OMM1_tau_vs_tx_oppo_and_same} (a,b)  for the opposite ($t^{\chi}_x = - t^{-\chi}_x $)  and same ($t^{\chi}_x =  t^{-\chi}_x $) tilt orientation, respectively, with positive (red) and negative (green) chiralities. This confirms the variation of the $\tau^{\chi}(\theta,\phi)$ with tilts at a fixed value of $\theta$ and $\phi$, naturally coming from the generalized momentum-dependent scattering time calculation using Fermi's golden rule}.



{
\section{Non-monotonic dependence of magneto-optical response on tilts $t_{x/z}$}
\label{App:Non-monotonic dependence of MOR on tilts $t_{x/z}$}


To get more insight into the non-monotonic dependence, we decompose the conductivity as follows:
\begin{align}
        \sigma_{zz}  &=\frac{-e^2}{\hbar} \sum_{\chi}\int\frac{d^3\mathbf{k}}{(2\pi)^3} \bigg[ \left(  ( \mathbf{v}_\mathbf{k}^\chi \cdot \boldsymbol{\Omega}^\chi_{\mathbf{k}})~ B \sin\gamma + \frac{\hbar}{e} v_{z,\textbf{k}}^{\chi}\right)  \nonumber \\ &\times ~\Lambda^{\chi}_{z, {\bf k}}\left( \frac{\partial f_{0}^{\chi}}{\partial \epsilon_{\bf{k}}^\chi}\right) \bigg]\nonumber \\
        &= \frac{-e^2}{\hbar} \sum_{\chi}\int\frac{d^3\mathbf{k}}{(2\pi)^3} \bigg[ \left(  ( \mathbf{v}_\mathbf{k}^\chi \cdot \boldsymbol{\Omega}^\chi_{\mathbf{k}})~ B \sin\gamma + \frac{\hbar}{e} v_{z,\textbf{k}}^{\chi}\right)  \nonumber \\ &\times \big[d^{\chi}-h^{\chi}_{z,\mathbf{k}} + a^{\chi}\cos{\theta} +b^{\chi}\sin{\theta}\cos{\phi}+c^{\chi}\sin{\theta}\sin{\phi} \big] \nonumber \\ &\times{\widetilde{\tau}^{\chi}_{\textbf{k}}(\theta,\phi)}\left( \frac{\partial f_{0}^{\chi}}{\partial \epsilon_{\bf{k}}^\chi}\right) \bigg]
        \nonumber \\ 
        &=\sigma_{zz}^d + \sigma_{zz}^h + \sigma_{zz}^a + \sigma_{zz}^b + \sigma_{zz}^c,
\end{align}
with,
\begin{align}
\mathcal{T}[X^\chi_{\mathbf{k}}]
&\equiv
-\frac{e^2}{\hbar}
\sum_{\chi}
\int\frac{d^3\mathbf{k}}{(2\pi)^3}
\left[
B\sin\gamma\,
\bigl(\mathbf{v}_{\mathbf{k}}^\chi\cdot
\boldsymbol{\Omega}_{\mathbf{k}}^\chi\bigr)
+\frac{\hbar}{e}v_{z,\mathbf{k}}^\chi
\right]\nonumber \\ &\times
X^\chi_{\mathbf{k}}\,
\widetilde{\tau}_{\mathbf{k}}^\chi(\theta,\phi)
\frac{\partial f_0^\chi}{\partial\epsilon_{\mathbf{k}}^\chi},
\label{eq:T-functional}
\\[2pt]
\sigma_{zz}^d&=\mathcal{T}[d^\chi], ~~~\sigma_{zz}^h=\mathcal{T}[-h_{z,\mathbf{k}}^\chi], ~~~\sigma_{zz}^a=\mathcal{T}[a^\chi\cos\theta], \nonumber \\ \sigma_{zz}^b &=\mathcal{T}[b^\chi\sin\theta\cos\phi],
~~~\sigma_{zz}^c=\mathcal{T}[c^\chi\sin\theta\sin\phi].
\label{eq:T-components}
\end{align}
Now in Figs.~\ref{fig:sigzz_OMM1_same_omgtau_lt_1_alp_vary} (a,b,c,d), we have plotted these individual contributions  $\delta\sigma^d_{zz}/\sigma(0)$, $\delta\sigma^h_{zz}/\sigma(0)$, $\delta\sigma^a_{zz}/\sigma(0)$, and $\delta\sigma^b_{zz}/\sigma(0)$, respectively, 
as a function of parallel tilt $t_{z}$ for the OMM \textit{on} case at $B=0.5~T$ and $\omega\tau\ll1$ regime. Note that the total contribution  $\delta\sigma_{zz}/\sigma(0)$ is presented in  Fig.~\ref{fig:LMC_vs_tz.png}(c). For a fixed value of $\alpha$, the total LMOC is governed by the term having the highest magnitude, e.g., for small $\alpha$ and tilt $t_z$, $\delta\sigma^d_{zz}/\sigma(0)$ dominates, and the overall nature is monotonic with tilt amplitude. Increasing $\alpha$ suppresses this contribution faster than the second leading contribution $\delta\sigma^h_{zz}/\sigma(0)$, which is non-monotonic in nature at higher values of tilts. Therefore, at a fixed value of $\alpha$, the competition between these two terms decides the trends while the effect of $\delta\sigma^{a,b,c}_{zz}/\sigma(0)$ is sub-leading. We would like to highlight that for a small tilt, the LMOC is monotonic, as studied in previous works using the constant-relaxation-time approximation where the  charge conservation is not guaranteed~\cite{gao2022suppression,gupta2024magneto,liu2021role,kundu2020magnetotransport,detassis2017collective,carbotte2016dirac,li2021nonlinear}. 

Therefore, the non-monotonic nature of LMOC is the direct consequence of the combination of momentum-dependent scattering time and the inclusion of the charge conservation constraint. For the regime $\omega\tau \gg1$, the presence of finite frequency overcomes the effect of $\alpha$, and shows non-monotonicity due to dominance of $\delta\sigma^h_{zz}/\sigma(0)$, which is not plotted explicitly.\\
For the sake of completeness, we extend our analysis to the tilt along the $x$-direction for the opposite tilt orientation. In Fig.~\ref{fig:tx_sigzz_OMM1_opp_omgtau_lt_1_alp_vary} (a,b,c,d), different contributions of LMOC i.e., $\delta\sigma^d_{zz}/\sigma(0)$, $\delta\sigma^h_{zz}/\sigma(0)$, $\delta\sigma^a_{zz}/\sigma(0)$, and $\delta\sigma^b_{zz}/\sigma(0)$, respectively, are plotted as a function of the perpendicular tilt $t_x$. For small tilts and weak intervalley scattering strength, the dominant contribution comes from $\delta\sigma^d_{zz}/\sigma(0)$, and the resultant LMOC is monotonic as plotted in Fig.~\ref{fig:LMC_vs_tx.png} (g). Increasing tilt, increases the contribution from $\delta\sigma^h_{zz}/\sigma(0)$ and dominates for $t_x \gtrsim 0.6 ~v_F$.  As a result, the total LMOC varies non-monotonically with the tilt amplitude in the case of  perpendicular tilt orientation as well. 
\begin{figure*}
    \centering
  \includegraphics[width=1.95\columnwidth]{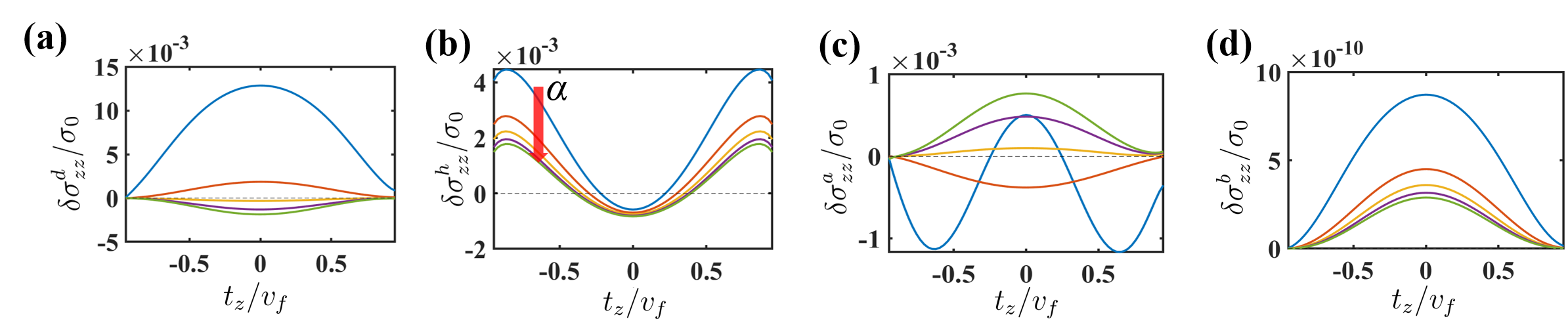}
    \caption{{Different components of LMOC as a function of tilt $t_z$ for parallel tilt orientation. (a-c) belong to $\omega\tau\ll1$. Along the arrow direction, blue to green, the intervalley scattering strength $\alpha$ increases from $0.2$ to $2.0$. We also, emphasze here that $\delta\sigma^c_{zz}/\sigma(0)$ is vanishingly small.}}
\label{fig:sigzz_OMM1_same_omgtau_lt_1_alp_vary}
\end{figure*}
\begin{figure*}
    \centering
  \includegraphics[width=1.95\columnwidth]{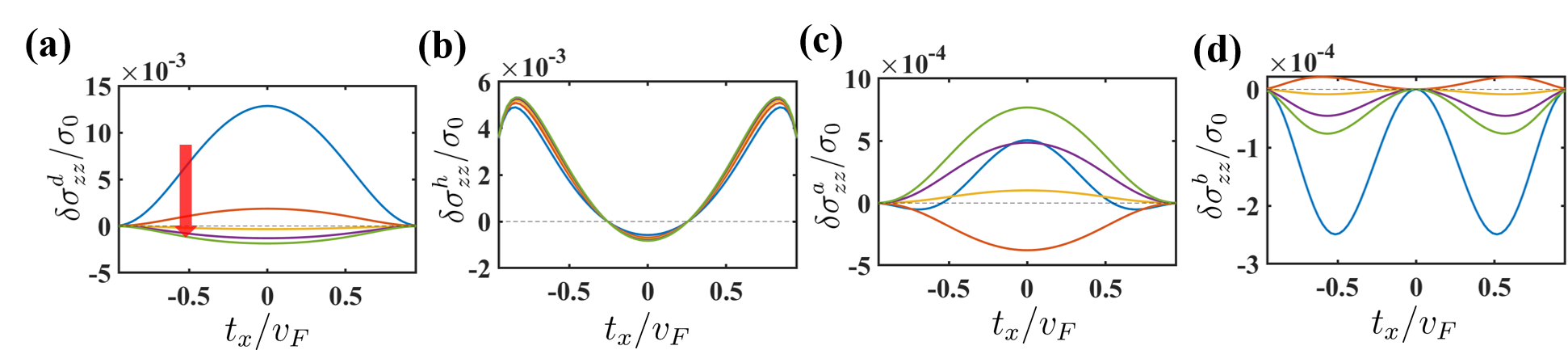}
    \caption{{Different components of LMOC as a function of tilt $t_x$ for opposite tilt orientation. (a-c) belong to $\omega\tau\ll1$. Along the arrow direction, blue to green, the intervalley scattering strength $\alpha$ increases from $0.2$ to $2.0$. We also, emphasze here that $\delta\sigma^c_{zz}/\sigma(0)$ is vanishingly small.}}
\label{fig:tx_sigzz_OMM1_opp_omgtau_lt_1_alp_vary}
\end{figure*}

}
\section{Imaginary part of magneto-optical conductivity responses}
\label{appc}

\begin{figure}
    \centering
    \includegraphics[width=.95\columnwidth]{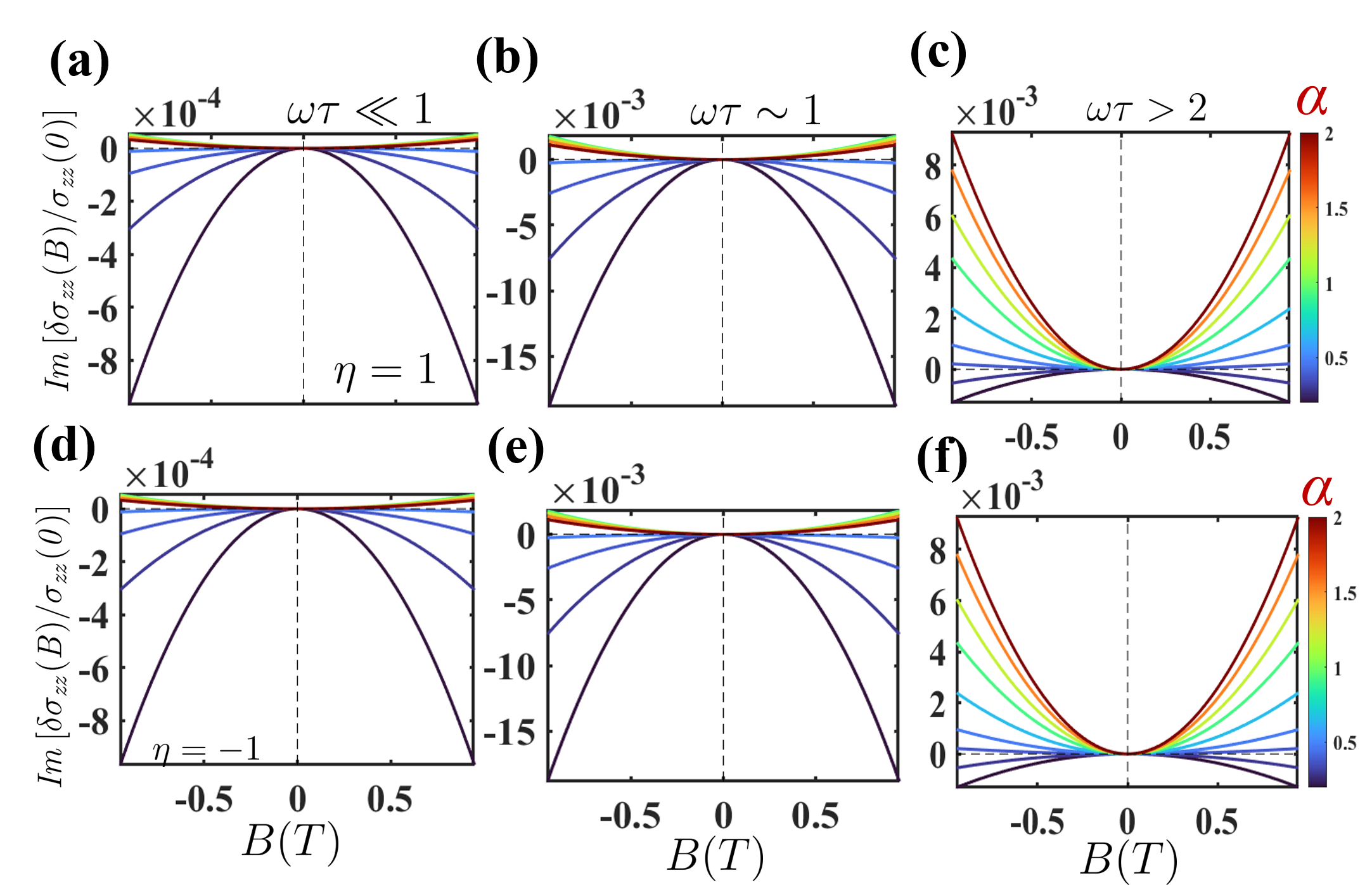}
    \caption{We repeat Fig. \ref{fig:Imag_LMC_vs_B_alp_vary1} for right and left circularly polarized light with $\eta=1$ and $-1$ in (a,b,c) and (d,e,f), respectively. Similar to the linearly polarized  light, the negative response even for strong intervalley scattering in weak ac limit, is a signature of the reactive part of LMOC.}
    \label{fig:Imag_LMC_vs_B_alp_vary1_eta_pm1}
\end{figure} 

The imaginary part of the longitudinal conductivity, $\mathrm{Im}\left[\sigma_{zz}(\omega)\right]$, originates directly from the complex definition of scattering time. For simplicity, we consider momentum-independent scattering time $\widetilde{\tau}^{\chi}_{\textbf{k}}(\theta,\phi)=\widetilde{\tau}^{\chi}_{\mu} $ which is independent of $\theta$ and $\phi$. The relaxation time under ac electric field is given by   $1/\widetilde{\tau}^{\chi}_{\mu} = \big(\frac{1}{\tau^{\chi}_{\mu}} -i\omega\big)$, which governs the frequency-dependent evolution of the nonequilibrium distribution function $f^{\chi}_\mathbf{k}$. We can rewrite the dressed relaxation time as follows  
\begin{align}
    \widetilde{\tau}^{\chi}_{\mu} = \frac{\tau^{\chi}_{\mu}}{1-i\omega\tau^{\chi}_{\mu}} =\frac{\tau^{\chi}_{\mu}(1+i\omega\tau^{\chi}_{\mu})}{1+(\omega\tau^{\chi}_{\mu})^2}= {\rm Re}[\widetilde{\tau}^{\chi}_{\mu}] + i~ {\rm Im}[\widetilde{\tau}^{\chi}_{\mu}],   
\end{align}
Note that the above breakdown also applies to the momentum-dependent scattering time $\widetilde{\tau}^{\chi}_{\textbf{k}}$ leading to the real and imaginary part of LMOC and TMOC, originated from ${\rm Re}[\widetilde{\tau}^{\chi}_{\textbf{k}}]$ and ${\rm Im}[\widetilde{\tau}^{\chi}_{\textbf{k}}]$, respectively.

It becomes clear that the reactive response arises from the $i\omega\tau^{\chi}_{\mu}$ term in the numerator, reflecting the finite time required for the accumulation  and relaxation of the chiral charge imbalance generated by the $\mathbf{E}\!\cdot\!\mathbf{B}$ term. While the real part of the conductivity describes dissipative energy absorption from the radiation field, the imaginary component encodes the phase lag between the induced current and the oscillatory electric field~\cite{heidari2020chiral}. This contribution becomes most pronounced in the intermediate regime $\omega \tau^{\chi}_{\mu}\sim1$, where the system neither relaxes completely within a single optical cycle nor follows the external field adiabatically. \\

We here show the imaginary part of the quantity $\delta \sigma_{\alpha\beta}(B)/\sigma_{\alpha\beta}(0)={\rm Im} [\sigma_{\alpha\beta}(B\ne 0)/\sigma_{\alpha\beta}(0)-1]$ by discarding the Drude contribution in
Figs.~\ref{fig:Imag_LMC_vs_B_alp_vary1} and ~\ref{fig:Imag_LMC_vs_B_alp_vary1_eta_pm1}. It can be clearly seen that as the strength of intervalley scattering $\alpha$ varies, both real and imaginary part of the conductivity shows sign reversal but in opposite way. Beyond a critical value of $\alpha = \alpha_c$, $\mathrm{Im}\,[\sigma_{zz}]$ exhibits a sign reversal, from negative to positive, signaling a qualitative change in the phase relation between the anomaly-driven current and the applied electric field. Physically, this sign change indicates that the reactive anomaly contribution overtakes the conventional transport component and reverses the direction of the phase-shifted current relative to the driving field.

\section{Effect of OMM on the angular profile of conductivities}
\label{appd}
\begin{figure}[ht]
    \centering
    \includegraphics[width=.95\columnwidth]{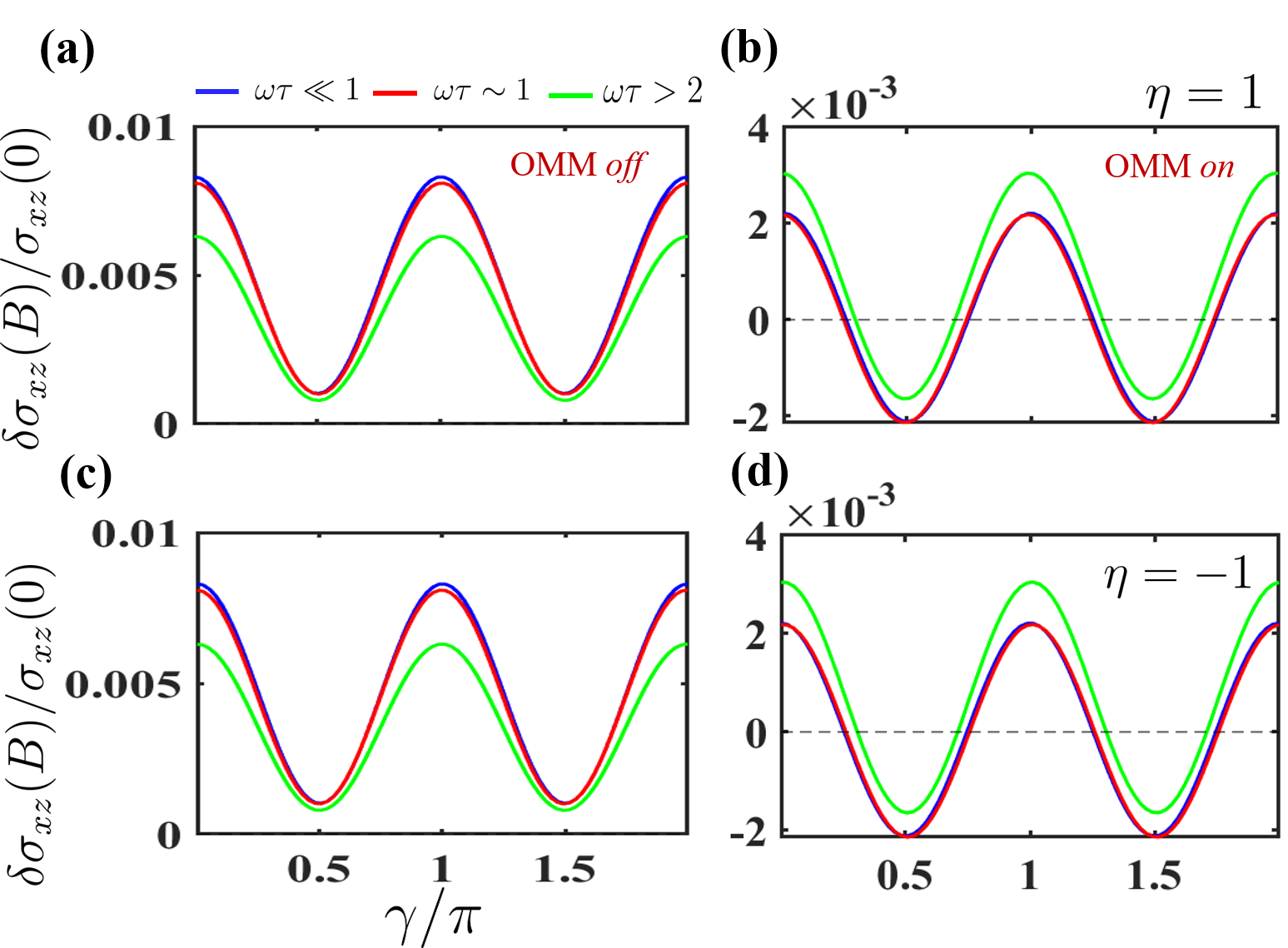}
    \caption{We show the angular variation of the real part of  TMOC in (a,b) and (c,d), under the application of right circularly and left circularly polarized light, respectively. In (a,c) [(b,d)], the OMM is off [on].  We illustrate the response in three different regimes $\omega \tau \ll1$, $\sim 1$, and $> 2$ with blue, red, and green lines, respectively. The static magnetic field $(B\cos \gamma, 0, B \sin\gamma)$ and electric field $(E_x,0,E_z)$, associated with the electromagnetic radiation, lie on the $xz$-plane causing the 
     $\sin 2\gamma$ response. The correction in linear-$B$, caused by OMM, results in the shift of TMOC with respect to the $\delta \sigma_{zx}(B)/\sigma_{zx}(0)=0$. We consider $\alpha=0.50$.}
    \label{fig:phc_vs_gm.png}
\end{figure}

\begin{figure}[ht]
    \centering
    \includegraphics[width=.95\columnwidth]{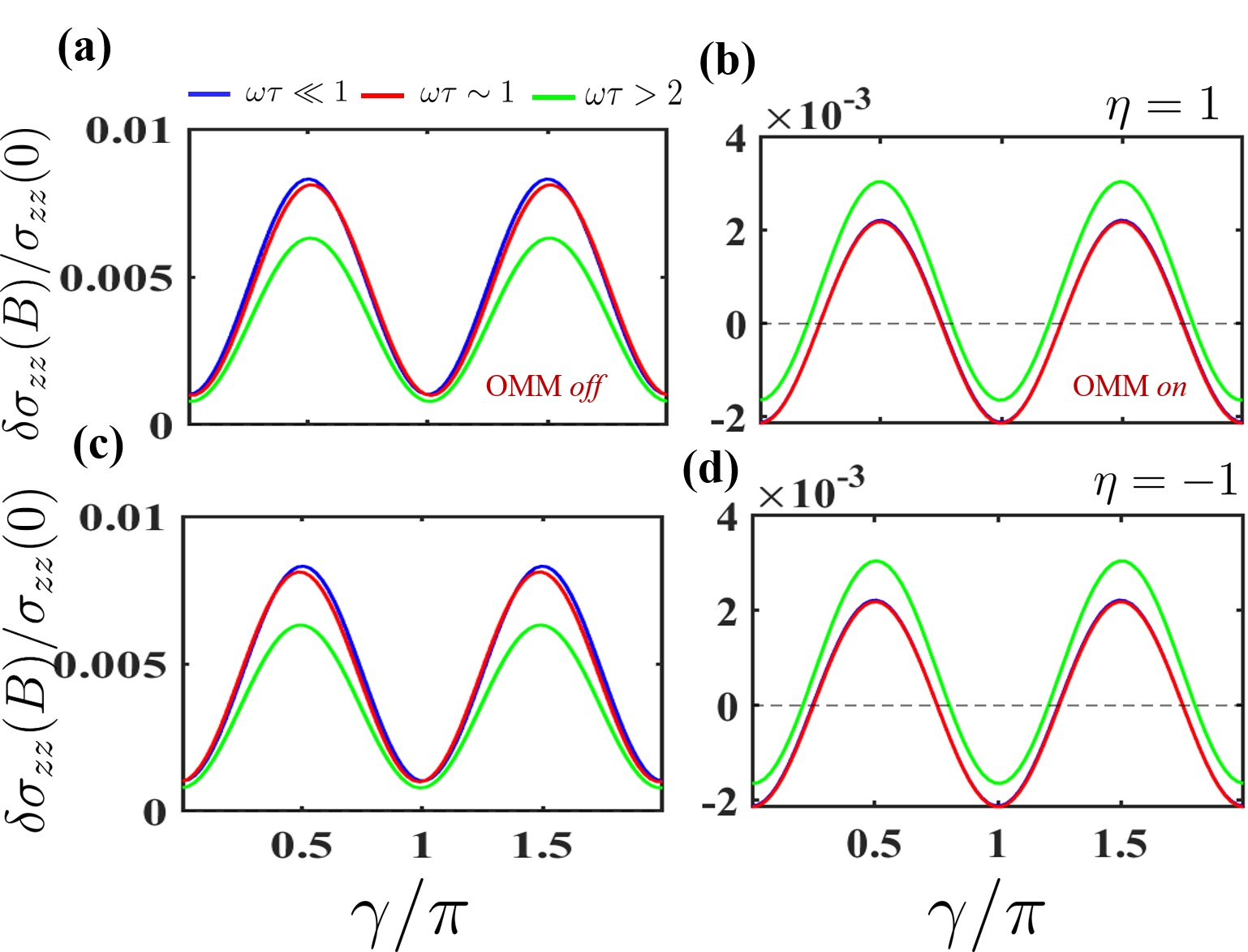}
    \caption{We repeat Fig. \ref{fig:phc_vs_gm.png} for the imaginary part of  LMOC in (a,b) and (c,d) under the application of right circularly and left circularly polarized light, respectively. The correction in linear-$B$, caused by OMM, results in the shift of LMOC with respect to the $\delta \sigma_{zz}(B)/\sigma_{zz}(0)=0$ while following $\sin^2 \gamma$ profile. We consider $\alpha=0.50$.} 
    \label{fig:LMC_vs_gm.png}
\end{figure}

To investigate the effect of the OMM on the angular dependence of LMOC and TMOC more extensively, we show their variation under circularly polarized light. In 
 Fig. \ref{fig:LMC_PHC_vs_gm.png} of the main text, we illustrate the shift in LMOC. In continuation, we demonstrate the shift in $\sin 2\gamma$ and $\sin^2 \gamma$ profiles for TMOC and LMOC in Figs. \ref {fig:phc_vs_gm.png}  and \ref{fig:LMC_vs_gm.png}, respectively. 


\end{document}